\documentclass[12pt]{iopart}
\usepackage{bm}
\usepackage{graphicx}
\begin{document}
\title{Keller's Theorem Revisited}
\author{Guillermo P. Ortiz$^1$ and W. Luis Moch\'an$^2$\\[5pt]
  $^1$Departamento de F{\'\i}sica,\\ Facultad de Ciencias Exactas
  Naturales y Agrimensura,\\
  Universidad Nacional del Nordeste,
  Corrientes, Argentina\\
  {\tt gortiz@unne.edu.ar}
  \\[5pt]
  $^2$Instituto de Ciencias F{\'\i}sicas,\\ Universidad Nacional
  Aut\'onoma de M\'exico,\\ Apartado Postal 48-3, 62251 Cuernavaca,
  Morelos, M\'exico\\ 
  {\tt mochan@fis.unam.mx}
}
\begin{abstract}
  Keller's theorem relates the
  components of the macroscopic dielectric response of a binary
  two-dimensional composite system with those of the reciprocal system
  obtained by 
  interchanging its components. We present a derivation of the theorem
  that, unlike previous ones, does not employ the 
  common asumption that the response function relates an irrotational
  to a solenoidal field and that is valid for dispersive and
  dissipative anisotropic systems. We show that the usual statement of
  Keller's theorem in terms of the conductivity is strictly valid only
  at zero frequency. We verify the theorem numerically in several
  ordered and disordered systems and discuss some of its consequences. 
\end{abstract}
\maketitle

\section{Introduction}\label{interoduccion}

In 1964, J. B. Keller ~\cite{Keller(1964)} showed that for binary
periodic composites made of particles in the shape of
generalized cylinders with arbitrary cross sections but with certain
mirror symmetries arranged in a
2D rectangular lattice within a host, the
macroscopic conductivity along a principal direction is
proportional to the inverse of the conductivity along the orthogonal
direction of the reciprocal system, obtained from the original system
by interchanging its constituent materials. The proportionality constant is the
product of the  conductivities of both materials.
This result, known as Keller's theorem,
was originally obtained by averaging the
microscopic current along an edge of the unit cell~\cite{Keller(1964)}
and writing it in terms of the electric potential, which is a solution
of Laplace's equation.

The conditions under which Keller's result applies
were later generalized, special cases were
discussed and some applications have been developed. Keller
\cite{Keller(1964)} showed that for a checkerboard 
geometry one could obtain a simple analytical formula for the macroscopic
conductivity as a simple consequence of his theorem: the
macroscopic response is given simply by the geometrical
mean of the conductivities of its two phases.
The same formula was then shown to apply to the  
conductivity of a macroscopically homogeneous and isotropic but
microscopically disordered 2D system made up of two phases with the
same total area \cite{Dykhne(1971)}.
From this formulae, approximate \cite{keller_effective_1987}
results for the conductivity of a 2D lattice of parallelograms and of
3D parallelepipeds for systems with high contrast have been found. 
Similar closed formulae have been proposed \cite{Mortola(1985)} and
proved \cite{Milton(2001)} for 2D checkerboard with more than two
phases. 

On the other hand, Keller's theorem has been
generalized \cite{Mendelson(1975)} to anisotropic 2D composites and
a relation has been found relating the tensors that describe the
macroscopic anisotropic response of a system to those of its
reciprocal, in which the microscopic responses are not only
interchanged but also rotated by a right angle. As a 
special case, the relation between the principal conductivities of
systems with isotropic components but anisotropic macroscopic response
were obtained~\cite{Mendelson(1975)}.

Schulgasser~\cite{Schulgasser(1976)} argued that a theorem analogous
to Keller's theorem, in which there is a unique
correspondence between the response of a system and that of its
reciprocal system cannot hold in 3D. He further provided a
counterexample consisting of an isotropic polycristalline system built
from a disordered mixture of randomly oriented anisotropic binary
layered crystallites . 
Molyneux \cite{Molyneux(1970)} has shown that for a disordered
homogeneous 3D system with components described by positive definite
tensors characterized by a stochastic functions with given one-, two-
and three-point
correlation functions one can establish strict bounds on the effective
permittivity but they cannot be improved on by incorporating further
correlations. For isotropic
biphasic tridimensional system it has been shown that the product of
the principal values of the macroscopic conductivity is bounded from
below by the product of the microscopic conductivities of the constitutive
phases~\cite{Schulgasser(1976),Keller(1985)}.

Keller's theorem can be adapted to all kinds of problems described by
similar equations. Though first derived for the electrical
conductivity, it also applies to the dielectric permittivity or the
thermal conductivity~\cite{Schulgasser(1992)}.
A recurring theme present in the derivations of Keller's theorem is that a
system is excited by an irrotational field, such as an electrostatic field, or
a thermal gradient, and the system responds by establishing a
solenoidal field, such as an electric current, a displacement field or
a heat flux. Then, use is made of the fact that a $\pi/2$ rotation
interchanges the irrotational and solenoidal character of a field in
2D, so that a rotated excitation (response) may be interpreted as the
response (excitation) for the reciprocal system. Thus, a question that
naturally arises concerns the possible generalization of Keller's
theorem to situations in which the excitation and response fields can
have a different
nature. For instance, the displacement field is solenoidal in the
absence of external charge, but Keller's theorem might be
applicable even in the presence of external charge. Similarly, an
electic current is necessarily solenoidal only in the stationary case,
but it is not so in the dynamical case, when excited by a time varying
field.  

The homogenization problem of a composite
excited by oscillating sources has been analyzed by Wellander using
the notion of two-scale convergence~\cite{Wellander(2003)} for systems
that occupy a finite region and when the sources of the excitation lie on
its outside. Guenneau
et. al. also generalized Keller's theorem to finite frequency
\cite{Guenneau(2007)}. 
An important physical limitation of the finite frequency
generalizations is the usual assumption that the system is characterized by
Hermitian response operators, thus 
excluding absorbing media~\cite{Guenneau(2007)}.
Some other approaches for the homogenization of Maxwell equations have been
proposed \cite{Halevi(1999),Krokhin(2002),Perez(2006),Silveirinha(2007)}. In 1985
Moch\'an and Barrera developed a  
general homogenization theory in term of projection operators that allow
accounting for the effects of the fluctuations of the microscopic
electromagnetic fields in the the macroscopic electromagnetic
response~\cite{Mochan(1985)I}. 
They also developed several applications of that homogenization
formalism to diverse systems such as liquids, bulk crystals,
crystalline surfaces and rough surfaces~\cite{Mochan(1985)II}. 
In this work we apply this formalism to extend Keller's theorem to
the dielectric response of a 2D binary composite in the finite
frequency case, allowing for dispersion and absorption, though we
remain in the non-retarded regime, where the wavelength of light is
assumed to be much larger than the lengthscale corresponding to the
microscopic texture of the material.

The paper is organized as
follows: In section \ref{teoria} we obtain Keller's theorem for the
dielectric tensor of 2D binary composites and study some special
cases, such as
isotropic systems and systems symmetric under interchange of
materials. We also obtain a finite frequency generalization of
Keller's theorem for the electrical conductivity. In section
\ref{aplicaciones} we develop some applications of the theory. Namely,
we show that the normal and parallel response functions of a
superlattice are determined one from the other; we test the compliance
of effective medium theories to Keller's condition; we test 
the accuracy of an efficient computational scheme based on Haydock's
recursive method calculation
\cite{Cortes(2010),Mochan(2010),Ortiz(2014)} for the 
calculation of the macroscopic response of periodic systems; we
discuss the relation among the dielectric resonances of a system and
that of its reciprocal system and we explore the corresponding
microscopic fields\cite{Toranzos(2017)}; we test the accuracy of
numerical computations for 
ensemble members of disordered systems; and we illustrate how Keller's
theorem may be used to increase the accuracy of rough approximate
theories. Finally, section \ref{conclusions} is devoted to
conclusions. In an appendix we generalize our results for the case of
a composite made of anisotropic components.

\section{Theory}\label{teoria}
Within a composite medium the electromagnetic fields have spatial
variations due to 
the finite wavelength of light. They also have spatial variations due to
the texture of the system. The macroscopic field has only the former
variations and we will treat the latter as spatial fluctuations which
we proceed to eliminate to obtain the macroscopic response
$\hat\epsilon_M$ of the system from its {\em microscopic} response
$\hat\epsilon$. 
The microscopic dielectric response $\hat\epsilon$ of a composite
media is in general a linear operator which acting on the microscopic
electric field $\vec E$ yields the displacement field
\begin{equation}\label{DeE}
  \vec D=\hat\epsilon \vec E,
\end{equation}
and it can be written as
\begin{equation}\label{matriz}
  \hat\epsilon=\left(
  \begin{array}{cc}
    \hat\epsilon_{aa}&\hat\epsilon_{af}\\
    \hat\epsilon_{fa}&\hat\epsilon_{ff}
  \end{array}
\right),  
\end{equation}
where we define
\begin{equation}\label{proyecciones}
  \hat\epsilon_{\alpha\beta}\equiv\hat\mathcal
    P_\alpha\hat\epsilon\hat\mathcal P_\beta, \quad\alpha,\beta=a,f,
\end{equation}
with $\hat\mathcal P_\alpha$ the average ($\alpha=a$) and the fluctuation
($\alpha=f$) projectors, defined such that for any field $\phi$,
$\phi_a\equiv\hat\mathcal P_a\phi$ is its average and $\phi_f=\hat\mathcal
P_f\phi$ its fluctuations around the average, so that Eq. (1) becomes
\begin{equation}\label{Da}
    \vec D_a=\hat\epsilon_{aa}\vec E_a+\hat\epsilon_{af} \vec E_f,  
\end{equation}
\begin{equation}\label{Df}
    \vec D_f=\hat\epsilon_{fa}\vec E_a+\hat\epsilon_{ff} \vec E_f.  
\end{equation}
We will not pursue at this point a specific definition of what we mean
by average and by fluctuation, but we demand that the corresponding
operators $\hat \mathcal P_\alpha$ are projectors into complementary
subspaces, that is, they should be idempotent, $\hat\mathcal
P_\alpha^2=\hat\mathcal P_\alpha$ ($\alpha=a,f$), their cross
products should be 
null, $\hat\mathcal P_a\hat\mathcal P_f=\hat\mathcal
P_f\hat\mathcal P_a=0$ and $\hat\mathcal P_a+\hat\mathcal P_f=\hat 1$
with $\hat 1$ the identity operator. This means that $\hat\mathcal
P_a$ throws the 
fluctuations away, so a second application leaves the result unchanged,
$\hat\mathcal P_f$ throws the average away, so that a second
application leaves the result unchanged, and throwing away the
fluctuations of a field from which the average has been eliminated
leaves nothing. We will also assume that these operators are space-
and time-invariant, so that they commute with spatial and temporal
derivatives. 

Assume we excite the system with {\em external} charges and currents
described by the densities $\rho$ and $\vec\jmath$ that have no
fluctuations, $\rho=\rho_a$, $\vec \jmath=\vec \jmath_a$, $\rho_f=0$, and $\vec
\jmath_f=0$. We may assume this conditions as, being external sources,
$\rho$ and $\vec j$ 
are unrelated to the texture of the composite. 
From Maxwell equations for monochromatic fields with frequency
$\omega=qc$ within non-magnetic media we obtain a
wave equation for the fluctuating electric field
\begin{equation}\label{wave}
  \frac{1}{q^2}\nabla\times\nabla\times \vec E_f=\vec
  D_f=\hat\epsilon_{fa}\vec E_a+\hat\epsilon_{ff}\vec E_f,
\end{equation}
which we formally solve for $\vec E_f$
\begin{equation}\label{Ef}
  \vec E_f=-\left(\left(\hat\epsilon+\frac{\nabla^2}{q^2}\hat\mathcal
  P^T\right)_{ff}\right)^{-1} \hat\epsilon_{fa}\vec E_a,
\end{equation}
where we replaced $\nabla\times\nabla\times\to-\nabla^2\hat\mathcal
P^T$ and, using Helmholtz theorem, we introduced the transverse
projector $\hat\mathcal P^T$ and 
its complement, the longitudinal proyector $\hat\mathcal P^L$, so that
for any vector field $\vec F$, $\vec F^T\equiv\hat\mathcal P^T\vec F$
and $\vec F^L\equiv\hat\mathcal P^L\vec F$ are its transverse and
longitudinal projections, obeying $\vec F=\vec F^T+\vec F^L$,
$\nabla\times \vec F^T=\nabla\times \vec F$,  $\nabla
\cdot\vec F^L=\nabla\cdot \vec F$, $\nabla
\cdot\vec F^T=0$, and $\nabla\times \vec F^L=0$. As expected,
$(\hat\mathcal P^\gamma)^2=\hat\mathcal P^\gamma$ ($\gamma=L,T$),
$\hat\mathcal P^L\hat\mathcal P^T=\hat\mathcal P^T\hat\mathcal P^L=0$,
and $\hat\mathcal P^T+\hat\mathcal P^L=\hat 1$. In Eq. (\ref{Ef}) we
denote by $((\ldots)_{ff})^{-1}$ the inverse of the operator
$(\ldots)$ {\em after} having restricted it to fluctuating fields.
Substituting
Eq. (\ref{Ef}) into (\ref{Da}) we obtain
\begin{equation}\label{DavsEa}
  \vec D_a=\left(\hat\epsilon_{aa} - \hat\epsilon_{af} \left(\left(\hat\epsilon
  + \frac{\nabla^2}{q^2}\hat\mathcal
  P^T\right)_{ff}\right)^{-1} \hat\epsilon_{fa}\right)\vec E_a=\hat\epsilon_M\vec E_a,
\end{equation}
where we identified the macroscopic dielectric response
\begin{equation}\label{epsM}
  \hat\epsilon_M=\hat\epsilon_{aa} - \hat\epsilon_{af} \left(\left(\hat\epsilon
  + \frac{\nabla^2}{q^2}\hat\mathcal
  P^T\right)_{ff}\right)^{-1} \hat\epsilon_{fa},
\end{equation}
as that which relates the average displacement to the average electric
field.

In analogy to Eqs. (\ref{Da}) and (\ref{Df}), we write
\begin{equation}\label{EEa}
    \vec E_a=\hat\epsilon^{-1}_{aa}\vec D_a+\hat\epsilon^{-1}_{af} \vec D_f,  
\end{equation}
\begin{equation}\label{EEf}
    \vec E_f=\hat\epsilon^{-1}_{fa}\vec D_a+\hat\epsilon^{-1}_{ff} \vec D_f,  
\end{equation}
where $\hat \epsilon^{-1}$ is the inverse dielectric operator. From
Maxwell equations we obtain a wave equation for the fluctuating
displacement field
\begin{equation}\label{waveD}
  \nabla^2\hat\mathcal P^T(\hat\epsilon^{-1}_{ff}\vec D_f
  +\hat\epsilon^{-1}_{fa}\vec D_a) = -q^2 \hat\mathcal P^T D_f,
\end{equation}
where we used the absence of fluctuating external charges $\rho_f=0$.
We solve this equation for $\vec D_f$ as
\begin{equation}\label{DDf}
  \vec
  D_f=-\left((\hat\epsilon^{-1}+q^2\nabla^{-2})_{ff}^{TT})\right)^{-1}
  \hat\epsilon^{-1}_{fa} \vec D_a, 
\end{equation}
 where we denote by $((\ldots)_{ff}^{TT})^{-1}$ the inverse of the
 operator $(\ldots)$ after restricting it to fluctuating transverse
 fields. Here we introduced the inverse Laplacian $\nabla^{-2}$ as a
 way to denote the Green's operator $\nabla^{-2}=\hat G$ for Poisson's equation,
 $\nabla^2\hat G=\hat 1$.  Substituting Eq. (\ref{DDf}) into
 (\ref{EEa}) we obtain 
 \begin{equation}\label{EavsDa}
   \vec E_a = \left(\hat\epsilon^{-1}_{aa} - \hat\epsilon^{-1}_{af}
   \left((\hat\epsilon^{-1}+q^2\nabla^{-2})_{ff}^{TT})\right)^{-1}
   \hat\epsilon^{-1}_{fa}\right) \vec D_a=\hat \epsilon^{-1}_M\vec D_a,   
 \end{equation}
 where we identified the macroscopic inverse dielectric response
 \begin{equation}\label{eps-1M}
   \hat\epsilon^{-1}_M = \hat\epsilon^{-1}_{aa} - \hat\epsilon^{-1}_{af}
   \left((\hat\epsilon^{-1}+ q^2\nabla^{-2})_{ff}^{TT})\right)^{-1}
   \hat\epsilon^{-1}_{fa}.   
 \end{equation}

 Up to this point, our results (\ref{epsM}) and (\ref{eps-1M}) are
 completely general, as we have introduced no approximation in their
 derivation. Now we will consider the long-wavelength approximation,
 in which we assume that the wavelength $\lambda$ of a freely
 propagating wave of frequency $\omega$ is much larger than the
 lengthscale $\ell$ that corresponds to the texture of the composite,
 $\lambda\gg\ell$. We expect that $\nabla^2$ acting on a fluctuating field
 to be of order $1/\ell^2$. Thus, it may safely be assumed that
 $\hat\epsilon$  is negligible compared to $\nabla^2/q^2$ in
 Eq. (\ref{epsM}) except very close to a resonance or for metallic
 media at frequencies where the penetration depth is close to its
 minimum. However, $\nabla^2/q^2$ appears multiplied by $\hat\mathcal
 P^T$, so its effect is null when acting on longitudinal fields. Thus,
 we may approximate Eq. (\ref{epsM}) by
 \begin{equation}\label{epsMa}
   \hat\epsilon_M=\hat\epsilon_{aa} - \hat\epsilon_{af}
   (\hat\epsilon^{LL}_{ff})^{-1} \hat\epsilon_{fa}.   
 \end{equation}
 Similarly, we may neglect $q^2\nabla^{-2}$ acting on fluctuating fields
 when compared with $\epsilon^{-1}$ in Eq. (\ref{eps-1M}) and
 approximate it by
\begin{equation}\label{eps-1Ma}
   \hat\epsilon^{-1}_M=\hat\epsilon^{-1}_{aa} - \hat\epsilon^{-1}_{af}
   ((\hat\epsilon^{-1})^{TT}_{ff})^{-1} \hat\epsilon^{-1}_{fa}.   
 \end{equation}

Finally, we take the longitudinal projection of Eq. (\ref{epsMa}) and
the transverse projection of Eq. (\ref{eps-1Ma}), and we employ the
block matrix theorem to obtain
\begin{equation}\label{epsMLL}
  (\hat\epsilon^{LL}_M)^{-1} = ((\hat\epsilon^{LL})^{-1})_{aa}
\end{equation}
and
\begin{equation}\label{epsM-1TT}
  ((\hat\epsilon^{-1}_M)^{TT})^{-1} = (((\hat\epsilon^{-1})^{TT})^{-1})_{aa}.
\end{equation}
These are the main results of ref. \cite{Mochan(1985)I}.

Consider now the specific form for the transverse and longitudinal
projectors,
\begin{equation}\label{PL}
  \hat P^L=\nabla\nabla^{-2}\nabla\cdot
\end{equation}
and
\begin{equation}\label{PT}
  \hat P^T=-\nabla\times\nabla^{-2}\nabla\times, 
\end{equation}
so that for any vector field $\vec F$ we have
\begin{equation}\label{FL}
  \vec F^L=\nabla\nabla^{-2}\nabla\cdot \vec F,
\end{equation}
\begin{equation}\label{FT}
  \vec F^T=-\nabla\times\nabla^{-2}\nabla\times \vec F,
\end{equation}

In the particular case of 2 dimensions (2D), for fields along the
$X-Y$ plane depending only on $x$ and $y$, we can rewrite
Eq. (\ref{FT}) as
\begin{equation}\label{FTR}
  \vec F^T=\nabla_R\nabla^{-2}\nabla_R\cdot\vec F,
\end{equation}
where we represent $\nabla$ as the two dimensional vector operator
$(\partial/\partial_x, \partial/\partial_y)$, and $\nabla_R$ is the
same operator after a $90^\circ$ rotation
\begin{equation}\label{nablaR}
  \nabla_R=\left(\frac{\partial}{\partial y},-\frac{\partial}{\partial
    x}\right)=\mathbf R\cdot\nabla,
\end{equation}
with 
\begin{equation}\label{R}
  \mathbf R=\left(
  \begin{array}{cc}
    0&1\\-1&0
  \end{array}
  \right)
\end{equation}
the rotation matrix, which coincides with the 2D Levy-Civita
antisymmetric tensor. To avoid ambiguities in our notation and to
eliminate the need for the dot products above, we represent vectors as
column matrices and rewrite Eqs. (\ref{FL}) and (\ref{FTR}) as matrix
products,
\begin{equation}\label{FLM}
  \vec F^L=\nabla\nabla^{-2}\nabla^t\vec F,
\end{equation}
and
\begin{equation}\label{FTRM}
  \vec F^T=\nabla_R\nabla^{-2}\nabla_R^t\vec F,
\end{equation}
with the superscript $t$ denoting transpose.

We consider now a binary composite system made up of two isotropic
local materials $A$, $B$, with corresponding dielectric functions
$\epsilon_A$ and $\epsilon_B$, so that
\begin{equation}\label{epsB}
  \epsilon(\vec r)=\epsilon_A (1-B(\vec r)) + \epsilon_B B(\vec r),
\end{equation}
where $B(\vec  r)=0,1$ is the characteristic function which takes the
value 1 (0) in the regions occupied by material $B$ ($A$). Notice that
\begin{equation}\label{tildeEps}
  \epsilon^{-1}(\vec r)=\frac{\tilde\epsilon(\vec r)}{\epsilon_A\epsilon_B},  
\end{equation}
where
\begin{equation}\label{tildeepsr}
  \tilde\epsilon(\vec r)=\epsilon_B(1-B(\vec r))+\epsilon_A B(\vec r)
\end{equation}
corresponds to the same composite as $\epsilon(\vec r)$ but with
material $A$ interchanged with material $B$. Thus, we write
Eq. (\ref{epsM-1TT}) as
\begin{eqnarray}
  \label{epsM-3}
  ((\hat\epsilon^{-1}_M)^{TT})^{-1} &=&
   \epsilon_A\epsilon_B
   ((\hat{\tilde\epsilon}^{TT})^{-1})_{aa}\\
  \label{epsM-4}
   &=&\epsilon_A\epsilon_B
   ((\nabla_R\nabla^{-2}\nabla_R^t \hat{\tilde\epsilon}
   \nabla_R\nabla^{-2}\nabla_R^t )^{-1})_{aa}\\ 
  \label{epsM-5}
   &=&\epsilon_A\epsilon_B
   ((\mathbf R \nabla\nabla^{-2}\nabla^t\mathbf R^t
   \,\hat{\tilde\epsilon}\,  \mathbf R
   \nabla\nabla^{-2}\nabla^t\mathbf R^t )^{-1})_{aa}, 
\end{eqnarray}
where we employed the transverse projector (Eq. (\ref{FTR}))
and introduced explicitly the 
rotation matrix $\mathbf R$ and its transpose $\mathbf R^t$. As
we assumed the microscopic response $\tilde\epsilon(\vec r)$ is
isotropic at each position, we can 
eliminate the innermost rotation matrices and write
\begin{eqnarray}
  \label{epsM-6}
  ((\hat\epsilon^{-1}_M)^{TT})^{-1} &=&\epsilon_A\epsilon_B
   ((\mathbf R \nabla\nabla^{-2}\nabla^t
   \hat{\tilde\epsilon}
   \nabla\nabla^{-2}\nabla^t\mathbf R^t )^{-1})_{aa} \\
   &=&
  \label{epsM-7}
   \epsilon_A\epsilon_B
   ((\mathbf R
   \hat{\tilde\epsilon}^{LL}\mathbf R^t )^{-1})_{aa} \\
      &=&
   \label{epsM-8}
   \epsilon_A\epsilon_B
   \mathbf R((
   \hat{\tilde\epsilon}^{LL})^{-1})_{aa}\mathbf R^t \\
   &=&
   \label{epsM-9}
   \epsilon_A\epsilon_B
   \mathbf R(
   \hat{\tilde\epsilon}_M^{LL})^{-1}\mathbf R^t, 
\end{eqnarray}
where we identified the longitudinal projector $\hat\mathcal P^L$
from Eq. (\ref{PL}) and the macroscopic dielectric function from
Eq. (\ref{epsMLL}). We invert both sides to obtain
\begin{equation}\label{Keller}
  (\hat\epsilon^{-1}_M)^{TT} =\frac{\mathbf
  R\hat{\tilde\epsilon}_M^{LL} \mathbf R^t}{\epsilon_A\epsilon_B}.
\end{equation}

Now we assume that the homogenized macroscopic system is
translationally invariant, so that its electromagnetic normal modes
are plane waves. Let the unit vector $\hat k$ be the direction of the
wavevector of 
any of such modes, and $\hat k_R=\mathbf R\cdot \hat k$ the perpendicular
direction. Then, we may interpret Eq. (\ref{Keller}) as
\begin{equation}\label{Kellerk}
    \hat k_R\hat k_R^t \boldsymbol\epsilon^{-1}_M \hat k_R\hat k_R^t =\frac{\mathbf
  R\hat k\hat k^t\tilde{\boldsymbol\epsilon}_M\hat k\hat
  k^t\mathbf R^t}{\epsilon_A\epsilon_B}, 
\end{equation}
where we introduced the representations $\hat\mathcal P^L\to\hat k\hat
k^t$ and  $\hat\mathcal P^T\to\hat k_R\hat
k_R^t$ of the longitudinal and transverse projectors in reciprocal
space, and we represent the dielectric operators $\hat\epsilon_M$ and
$\hat{\tilde\epsilon}_M$ by 
the dielectric tensors $\boldsymbol\epsilon_M$ and
$\tilde{\boldsymbol\epsilon}_M$. Introducing explicitly 
the rotation matrices, we rewrite this 
equation as 
\begin{equation}\label{KellerRk}
    \mathbf R\hat k\hat k^t \mathbf R^t
    \boldsymbol\epsilon^{-1}_M \mathbf R \hat k\hat k^t\mathbf R^t
    =\frac{\mathbf 
  R\hat k\hat k^t\tilde{\boldsymbol\epsilon}_M\hat k\hat
  k^t\mathbf R^t}{\epsilon_A\epsilon_B}. 
\end{equation}
We cancel the external rotation matrices, and since this
equation is obeyed for arbitrary directions $\hat k$, we also
cancel the projectors $\hat k\hat k^t$ to obtain finally our main
result, a version of Keller's interchange theorem
\begin{equation}\label{KellerF}
  \boldsymbol\epsilon_M \tilde{\boldsymbol\epsilon}_{MR} =
  \epsilon_A\epsilon_B\mathbf 1, 
\end{equation}
i.e., the macroscopic dielectric tensor of a binary composite
$\boldsymbol\epsilon_M$ multiplied by the rotated macroscopic dielectric
tensor of the same system but with the two materials
interchanged,
\begin{equation}\label{epsMR}
   \tilde{\boldsymbol\epsilon}_{MR} = \mathbf
   R\tilde{\boldsymbol\epsilon}_M \mathbf R^t, 
\end{equation}
is simply given by the product of the dielectric
functions of the components (times the identity tensor $\mathbf 1$).

We remark that to obtain this result we didn't assume the absence of
external charges nor currents. The response functions of the system
ought to be intrinsic quantities, with no dependence on the existence
of external sources. Actually, some homogenization theories require
external sources in their formulation. We only assumed
that the sources have no spatial fluctuations, as otherwise it
wouldn't make sense to pursue a macroscopic description of the
response of the system. Furthermore, we
made no assumption about the frequency, except for demanding that the
corresponding wavelength be large in comparison with the microscopic
lengthscale corresponding to 
the texture of the composite. The system may be periodic or random, as
we only demanded that from a macroscopic point of view it should be
homogeneous. The response functions of the components $\epsilon_A$ and
$\epsilon_B$ may be real positive constants, corresponding to
transparent dielectrics, or complex frequency dependent functions,
corresponding to dissipative, dispersive media.

Some simple consequences of Eq. (\ref{KellerF}) follow: The
determinant of Eq. (\ref{KellerF}) yields 
\begin{equation}\label{determinant}
  \det(\boldsymbol\epsilon_M) \det(\tilde{\boldsymbol\epsilon}_M) =
  \epsilon_A^2\epsilon_B^2.  
\end{equation}
In normal axes, say $X$, $Y$, it becomes
\begin{equation}\label{normalaxes}
  \epsilon_M^{xx}\tilde\epsilon_M^{yy} =
  \epsilon_M^{yy}\tilde\epsilon_M^{xx} = \epsilon_A\epsilon_B.   
\end{equation}
For isotropic (within the plane) composites it yields
\begin{equation}\label{isotropic}
  \epsilon_M\tilde\epsilon_M=\epsilon_A\epsilon_B,
\end{equation}
for the corresponding scalar response functions.
Finally, for the very special case of an isotropic system that is
invariant under the interchange $\epsilon_A\leftrightarrow\epsilon_B$,
such as a periodic checkerboard or a disordered system made by adding
randomly particles of each material with the same probability, we
obtain from Eq. (\ref{isotropic}) the analytical result
\begin{equation}\label{symmetric}
  \epsilon_M=\tilde\epsilon_M=\sqrt{\epsilon_A\epsilon_B}.
\end{equation}

We recall that the dielectric function $\epsilon_\alpha$, $\alpha=A,B$
of each phase may be written in terms of its conductivity
$\sigma_\alpha$ as
\begin{equation}\label{epssigma}
  \epsilon_\alpha=1+\frac{4\pi i\sigma_\alpha}{\omega},
\end{equation}
where we incorporate in $\sigma_\alpha$ the induced currents
within the system, including polarization and conduction
currents. Similarly, the macroscopic response may be written in terms
of a macroscopic conductivity,
\begin{equation}\label{epsMsigma}
  \boldsymbol\epsilon_M=1+\frac{4\pi i\boldsymbol\sigma_M}{\omega}.
\end{equation}
Subsititution of Eqs. (\ref{epssigma}) and (\ref{epsMsigma}) in
(\ref{KellerF}) yields
\begin{equation}\label{KellerSigma}
  \boldsymbol\sigma_M \tilde{\boldsymbol\sigma}_{MR}
  -\frac{i\omega}{4\pi}(\boldsymbol\sigma_M+\tilde{\boldsymbol\sigma}_{MR})
  = \sigma_A\sigma_B \mathbf 1
  -\frac{i\omega}{4\pi}(\sigma_A+\sigma_B)\mathbf 1,
\end{equation}
where we used a notation analogous to that in Eq. (\ref{epsMR}).
Thus, for low frequencies we recover the usual Keller's theorem for
the conductivity
\begin{equation}\label{KellerSigmaUsual}
  \boldsymbol\sigma_M \tilde{\boldsymbol\sigma}_{MR}
  = \sigma_A\sigma_B\mathbf 1,
\end{equation}
but this equality {\em is not obeyed at intermediate frequencies} and
at large frequencies it should be replaced by a new relation
\begin{equation}\label{KellerSigmaNuevo}
  \boldsymbol\sigma_M+\tilde{\boldsymbol\sigma}_{MR}
  =(\sigma_A+\sigma_B)\mathbf 1. 
\end{equation}
We remark that Keller's theorem was originally obtained for the
conductivity but assuming explicitly that the divergence
$\nabla\cdot\vec j=0$ of the electric
current density $\vec j$ is zero, and using that a $\pi/2$ rotation changes
curl-free fields to divergenceless fields and viceversa. However, that
derivation becomes invalid at finite frequencies, for which
$\nabla\cdot\vec j=i\omega\rho$ which in general is not null.

\section{Applications}\label{aplicaciones}

In this section we illustrate our generalized Keller's theorem with a
few applications and some numerical calculations.

\subsection{One dimensional systems}
Consider a 1D system made up by stacking thin layers of materials $A$
and $B$ along the $y$ direction. The electric $E^x$ parallel to the
layer surfaces is continuous across the interfaces and has a slow
spatial variation across an individual layer, so it is almost
constant. Thus, the macroscopic response
\begin{equation}\label{1Dxx}
  \epsilon_M^{xx}=(1-f)\epsilon_A+f\epsilon_B=\langle\epsilon\rangle
\end{equation}
is simply the average of the response of the components, where $f$ is
the filling fraction of the $b$ material. According to
Eq. (\ref{KellerF})
\begin{equation}\label{1Dyy}
  \frac{1}{\epsilon_M^{yy}} =
  \frac{\tilde\epsilon_M^{xx}}{\epsilon_A\epsilon_B} =
  \frac{(1-f)\epsilon_B+f\epsilon_A}{\epsilon_A\epsilon_B} =
  \frac{1-f}{\epsilon_A} + \frac{f}{\epsilon_B}=\left\langle\frac{1}{\epsilon}\right\rangle.
\end{equation}
This is a well known result which may be obtained by realizing that
$D^y$ is continuous across the interfaces and slowly varying across
each layer, so that the inverse dielectric function is the average of
the inverse dielectric functions of the components. Nevertheless,
we have shown that according to Keller's theorem the results above
are not independent, but 
{\em each one is a consequence of the other}.

\subsection{Effective medium theories}\label{sub:EMT}

In 2D Maxwell-Garnett theory assumes particles in the shape of
circular cylinders each of which responds to the local field, given by
an external 
field and the fields produced by all other particles, which is assumed
to be dipolar. Assuming the particles are on a square lattice or that
their positions are disordered but with no correlations beyond two
particle correlations, the field produced by particles within a
Lorentz cylindrical cavity would be null, while the field of those
particles farther away corresponds to the sum of the macroscopic field
and the depolarization field of the cavity, yielding the
expression\cite{Sihvola(1999)} 
\begin{equation}\label{MG}
  \frac{\epsilon_M-\epsilon_A}{\epsilon_M+\epsilon_A}=f
  \frac{\epsilon_B-\epsilon_A}{\epsilon_B+\epsilon_A}. 
\end{equation}
This formula equates the polarizability of a
cylinder made of the homogenized composite with the response
$\epsilon_M$ within a host with response $\epsilon_A$ with
the volume average 2D polarizability of cylinders with response
$\epsilon_B$ within the  
host $A$, i.e., the polarizability weighted by
the filling fraction $f$ of material $B$.

Interchanging materials
yields the response 
of the reciprocal system
\begin{equation}\label{MGt}
  \frac{\tilde\epsilon_M-\epsilon_B}{\tilde\epsilon_M+\epsilon_B}=f
  \frac{\epsilon_A-\epsilon_B}{\epsilon_A+\epsilon_B}.   
\end{equation}
As the right hand sides of equations (\ref{MG}) and (\ref{MGt}) are
equal but for a sign change, we may write
\begin{equation}\label{MGMGt}
  \frac{\tilde\epsilon_M-\epsilon_B}{\tilde\epsilon_M+\epsilon_B}=
  -\frac{\epsilon_M-\epsilon_A}{\epsilon_M+\epsilon_A},
\end{equation}
from which Eq. (\ref{isotropic}) follows immediately.

On the other hand, the symmetrical Bruggeman's effective medium theory
doesn't differentiate between host and particles and treats both
materials $A$ and $B$  on the same footing. It postulates that 
the average polarizability of particles made up of
materials $A$ and $B$ within a host made up of the homogenized
composite, weighted with the corresponding filling
fractions $1-f$ and $f$, should be null. For circular cylindrical
particles, this is represented by the equation\cite{Sihvola(1999)} 
\begin{equation}\label{bruggeman}
  (1-f)\frac{\epsilon_A-\epsilon_M}{\epsilon_A+\epsilon_M}+
  f\frac{\epsilon_B-\epsilon_M}{\epsilon_B+\epsilon_M}=0.
\end{equation}
We may rewrite this equation as
\begin{equation}\label{Br}
  \epsilon_M = \frac{\epsilon_A\epsilon_B}{\epsilon_M} + (1-2f)
  (\epsilon_A-\epsilon_B).
\end{equation}
When the media $A$ and $B$ are interchanged, this equation becomes
\begin{equation}\label{Brt}
  \tilde\epsilon_M = \frac{\epsilon_A\epsilon_B}{\tilde\epsilon_M} - (1-2f)
  (\epsilon_A-\epsilon_B).
\end{equation}
Adding Eqs. (\ref{Br}) and (\ref{Brt}) yields
\begin{equation}\label{BrAdd}
  \epsilon_M+\tilde\epsilon_M =
  \epsilon_A\epsilon_B\ \left(\frac{1}{\epsilon_M} +
  \frac{1}{\tilde\epsilon_M}\right), 
\end{equation}
from which Eq. (\ref{isotropic}) follows immediately.

\subsection{Periodic system}

\begin{figure}
  $$
  \includegraphics[width=.5\textwidth]{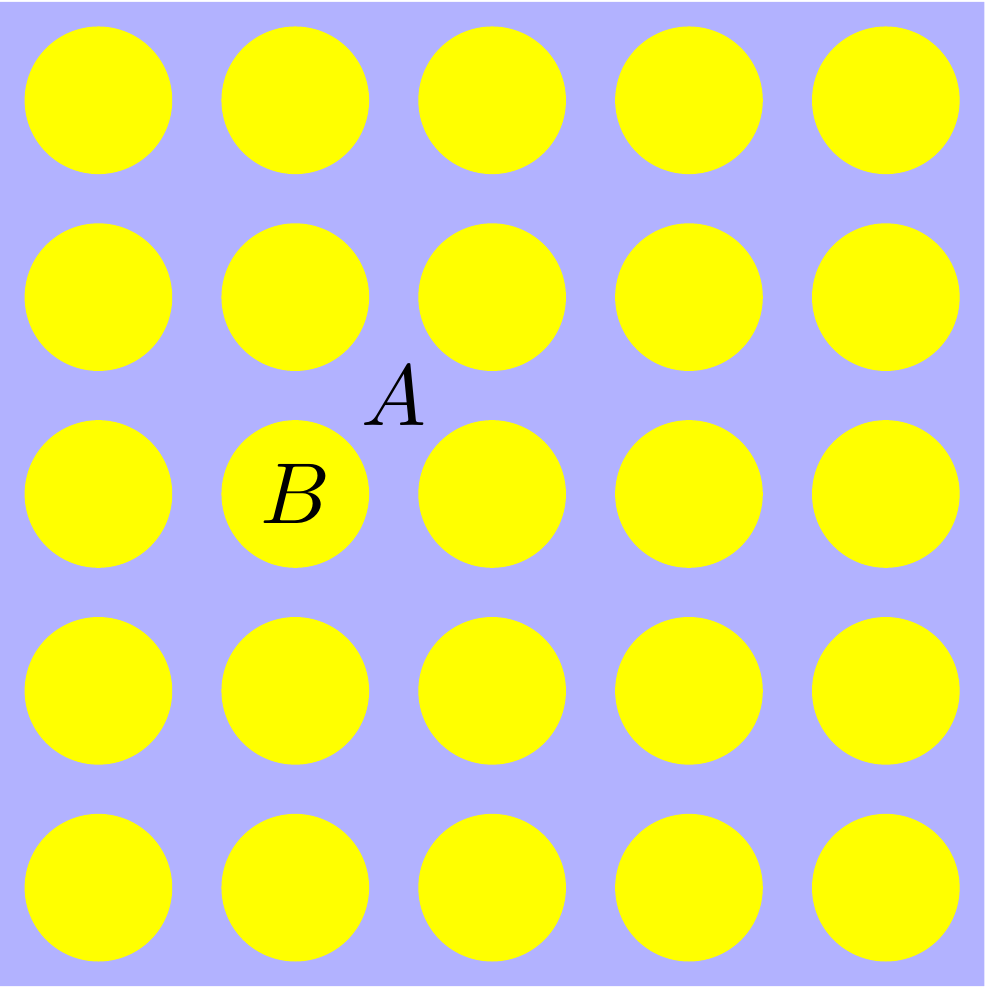}
  $$
  \caption{\label{fig:periodico} 
    Cross section of a square lattice of cylindrical inclusions $A$
    with response $\epsilon_A$ within a host $B$ of response
    $\epsilon_B$.
    }
\end{figure}

To illustrate the use of Keller's theorem to test numerical
calculations of the macroscopic dielectric response, we first consider
a square array of cylindrical metallic wires in vacuum and its
reciprocal system made up of a square array of cylindrical holes
within a metallic host (Fig. \ref{fig:periodico}). For simplicity we
model the metallic phase with the Drude response
\begin{equation}\label{ec:drude}
  \epsilon_D(\omega)=1-\omega_p^2/(\omega^2+i\omega\gamma)
\end{equation}
with a moderate damping characterized by the mean collision frequency
$\gamma=0.01\omega_p$. 
We calculate $\epsilon_M$ and $\tilde\epsilon_M$ for these systems
employing an efficient procedure
\cite{Ortiz(2009),Mochan(2010),Cortes(2010),Mendoza(2012),Huerta(2013)}
based on Haydock's recursive method (HRM) \cite{Haydock(1980)} and
implemented in the {\em Photonic} computational package
\cite{Mochan(2016)}. 

\begin{figure}
  \includegraphics[width=\textwidth]{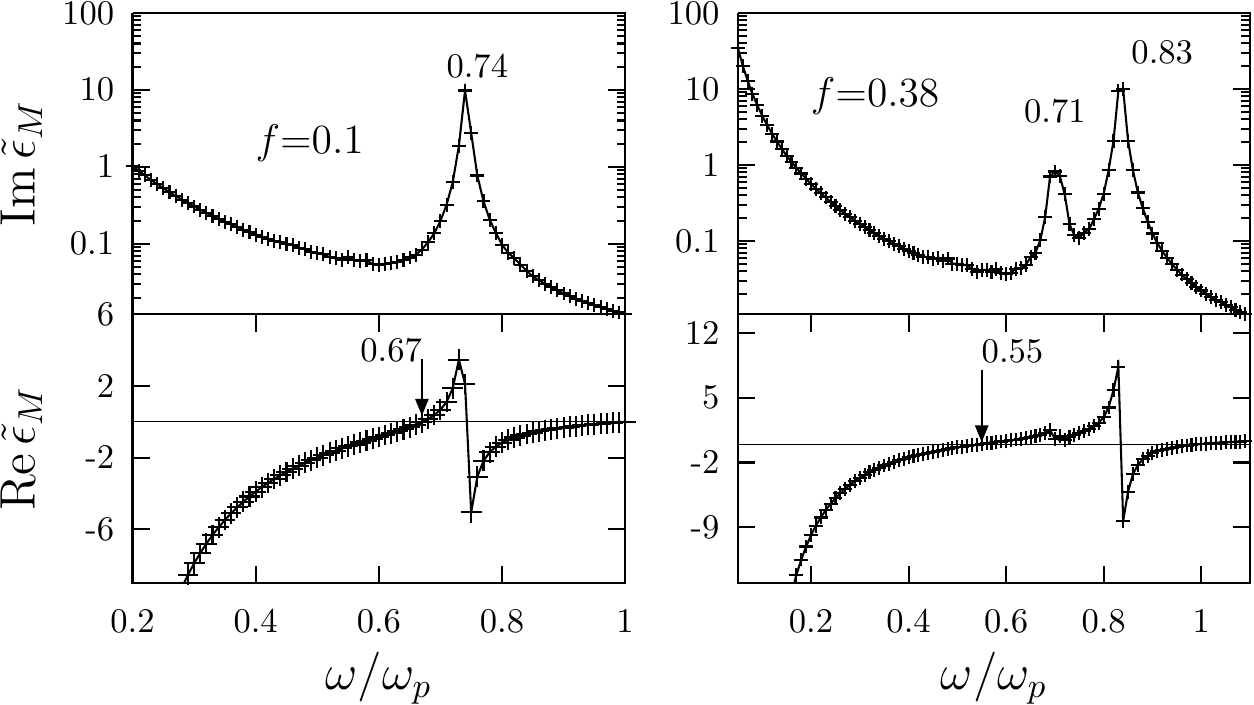}
  \caption{\label{fig:DiluOrd} Real and imaginary parts of the
    macroscopic response $\tilde\epsilon_M$ of a square lattice of
    cylindrical holes within a Drude metal as a function of frequency
    for a relatively low (left panels) and an
    intermediate (right panel) filling fraction $f=0.1$ and
    $f=0.38$. The continuous line corresponds
    to the HRM numerical calculation of $\tilde\epsilon_M$. The
    crosses correspond to the use of Keller's theorem to obtain
    $\tilde\epsilon_M$ from the response $\epsilon_M$ of an array of
    wires in vacuum, obtained from the HRM with the same Haydock
    coefficients but a different spectral variable. We indicate the
    frequencies of the peaks of $\mathrm{Im}\, \tilde\epsilon_M$ and
    the zeroes of $\mathrm{Re}\,\tilde\epsilon_M$.
  }
\end{figure}

In Fig. \ref{fig:DiluOrd} we show 
the response $\tilde\epsilon^M$  of an array of holes within a metallic
host with a small filling fraction $f=0.1$, calculated with the HRM. We
also show $\tilde\epsilon_M$ as obtained through the use of
Keller's theorem from the response $\epsilon_M$ of an array of wires
in vacuum, calculated with the HRM. The
agreement between both calculations is very good even at the
peaks.  In the figure we have indicated the resonance frequency
$\tilde\omega^{(1)}\approx0.74\omega_p$,
corresponding to a peak in $\mathrm{Im}\,\tilde\epsilon_M$. This resonance
is a dipolar resonance and is slightly blue shifted from that
corresponding to the dipolar surface plasmon of a single cylindrical
hole, at $\tilde\omega_d=\omega_p/\sqrt2$ 
due to the interaction with neighbor holes. We also indicate in the
figure the zero $\omega^{(1)}\approx0.67\omega_p$ of the real part of
$\mathrm{Re}\,\tilde\epsilon_M$, which, according to Keller's theorem,
corresponds to a resonance in the response $\epsilon_M$ of an array of
wires. This is slightly red-shifted with respect to the dipolar
surface plasmon $\omega_d=\omega_p/\sqrt2=\tilde\omega_d$ of a single
cylindrical 
wire.

In Fig. \ref{fig:DiluOrd} we also show results for a system with a
higher filling fraction $f=0.38$. The HRM
calculation for a lattice of holes and the application of Keller's
theorem to the HRM calculation for a lattice of wires are again in
very good agreement. In this case the interactions among inclusions
are stronger and the dipolar peak is further blue shifted up to
$\tilde\omega^{(1)}\approx0.83\omega_p$, while the zero is red shifted
to $\omega^{(1)}\approx0.55\omega_p$.

Notice that for both $f=0.1$ and $f=0.38$, the resonances
$\omega^{(1)}$ and $\tilde\omega^{(1)}$ are well described by the 2D
Maxwell-Garnett theory (Eqs. (\ref{MG}) and (\ref{MGt})), which for
this system yield $\omega^{(1)}=\sqrt{((1-f)/2)}\omega_p$ and
$\tilde\omega^{(1)}=\sqrt{((1+f)/2)}\omega_p$. Nevertheless, for
$f=0.38$ there is a further resonance at
$\omega^{l}\approx0.71\omega_p$. This is related to the excitation at
large filling fractions of multipoles of higher order than the
dipole. Curiously, for a cylindrical single wire and for a single hole
all the multipolar resonances are degenerate with the dipolar surface
plasmon at $\omega_p/\sqrt{2}$.

\begin{figure}
  $$
  \includegraphics[width=\textwidth]{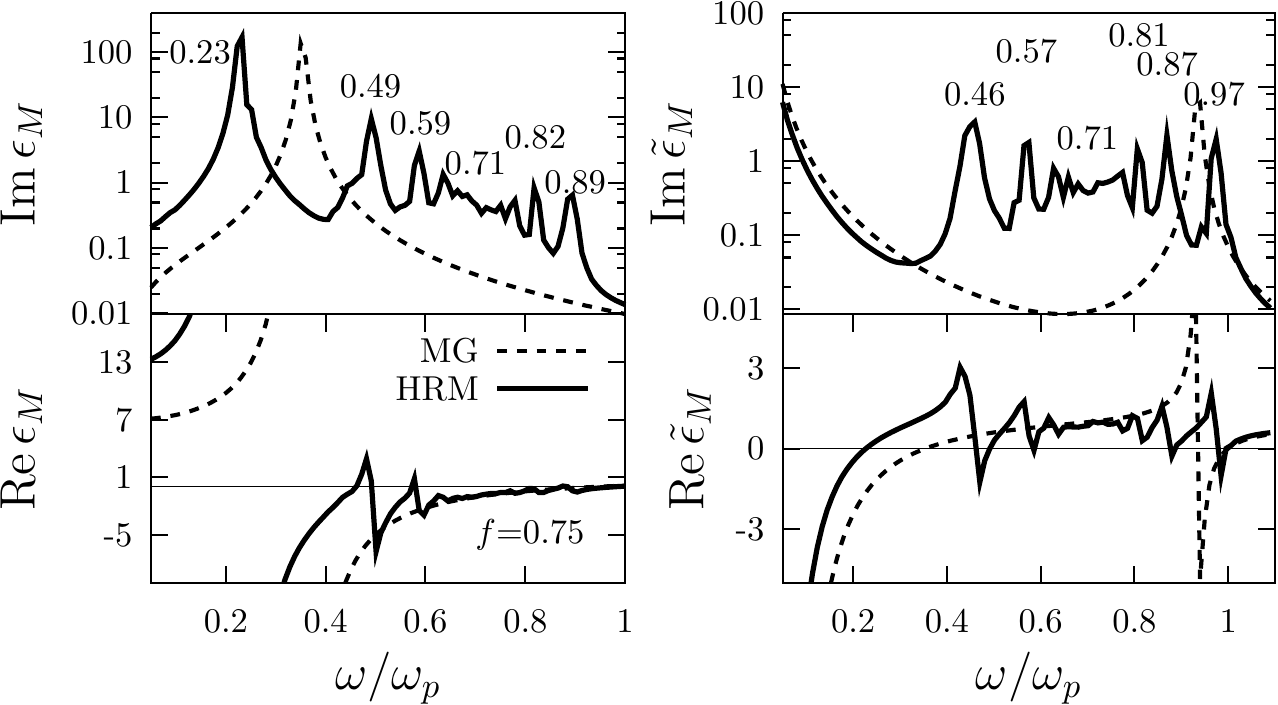}
  $$
  \caption{\label{fig:ConcentOrd} Real and imaginary parts of the
    macroscopic dielectric response of a square lattice of
    cylindrical wires in vacuum (right panels) and a square lattice
    of cylindrical holes within a metal for a high filling fraction
    $f=0.75$, calculated with the HRM 
    (solid) and with the MG approxaimation (dashed). 
    The conducting phases are described by the Drude
    response. The frequencies of a few resonances are indicated, as
    well as the resonance frequency of an isolated wire or hole
    close to $0.71\omega_p$.  
  }
\end{figure}

In Fig.\ref{fig:ConcentOrd} we show  $\epsilon^M$ and $\tilde
\epsilon^M$ calculated with the HRM for the same system as in
Fig. \ref{fig:DiluOrd} but with a high filling fraction $f=0.75$.
As a reference, we also show the results of MG
theory. While MG predicts a single peak with a dipolar character, the
numerical HRM calculation yields several peaks with multipolar
contributions, five of which are clearly visible. Of these, some are
blue shifted and some are red shifted with respect to the resonant
frequency of an isolated wire and an isolated hole. We expect that the
peak in $\epsilon_M$ that is furthest red shifted and the peak in
$\tilde\epsilon_M$ that is furthest blue shifted correspond to the
modes with the largest dipolar contribution. Both of these shifts are
close but larger than those predicted by MG theory.

The results above can be understood from the fact that within the
HRM we can write
\begin{equation}\label{Haydock}
\epsilon_M=\epsilon_A F(u),\quad \tilde\epsilon_M=\epsilon_B F(\tilde
u),
\end{equation}
where $u=1/(1-\epsilon_A/\epsilon_B)$ and $\tilde
u=1/(1-\epsilon_B/\epsilon_A)$ are the {\em spectral variables} of the
system and its reciprocal system, and where $F$ is a
function given by a continued fraction determined by the {\em
  Haydock} coefficients which are determined exclusively by the
geometry of the system. 
Notice that $\tilde u=1-u$, so that
any resonance $u^*$ in the function $F$ corresponds to a resonance
frequency $\omega^*$ in the system, such that $u(\omega^*)=u^*$, and a
corresponding resonance $\tilde\omega^*$ in the reciprocal system,
such that $\tilde u(\tilde\omega^*)=1-u(\tilde\omega^*)=u^*$. 
Thus, according to Keller's theorem, for each resonance $\omega_n$ in
$\epsilon_M$ there must be a corresponding resonance $\tilde\omega_n$
in $\tilde\epsilon_M$ and for the Drude model, they should be related
through $\omega_n^2+\tilde\omega_n^2=\omega_p^2$. From
Fig. \ref{fig:ConcentOrd} we can verify that this is the case
as
$0.23^2+0.97^2 \approx 0.49^2+0.87^2 \approx 0.59^2+0.81^2 \approx
0.82^2+0.57^2 \approx 0.89^2+0.46^2\approx1$.

\begin{figure}
  $$
  \includegraphics[trim = 25 30 50 70,width=0.3\textwidth]
                  {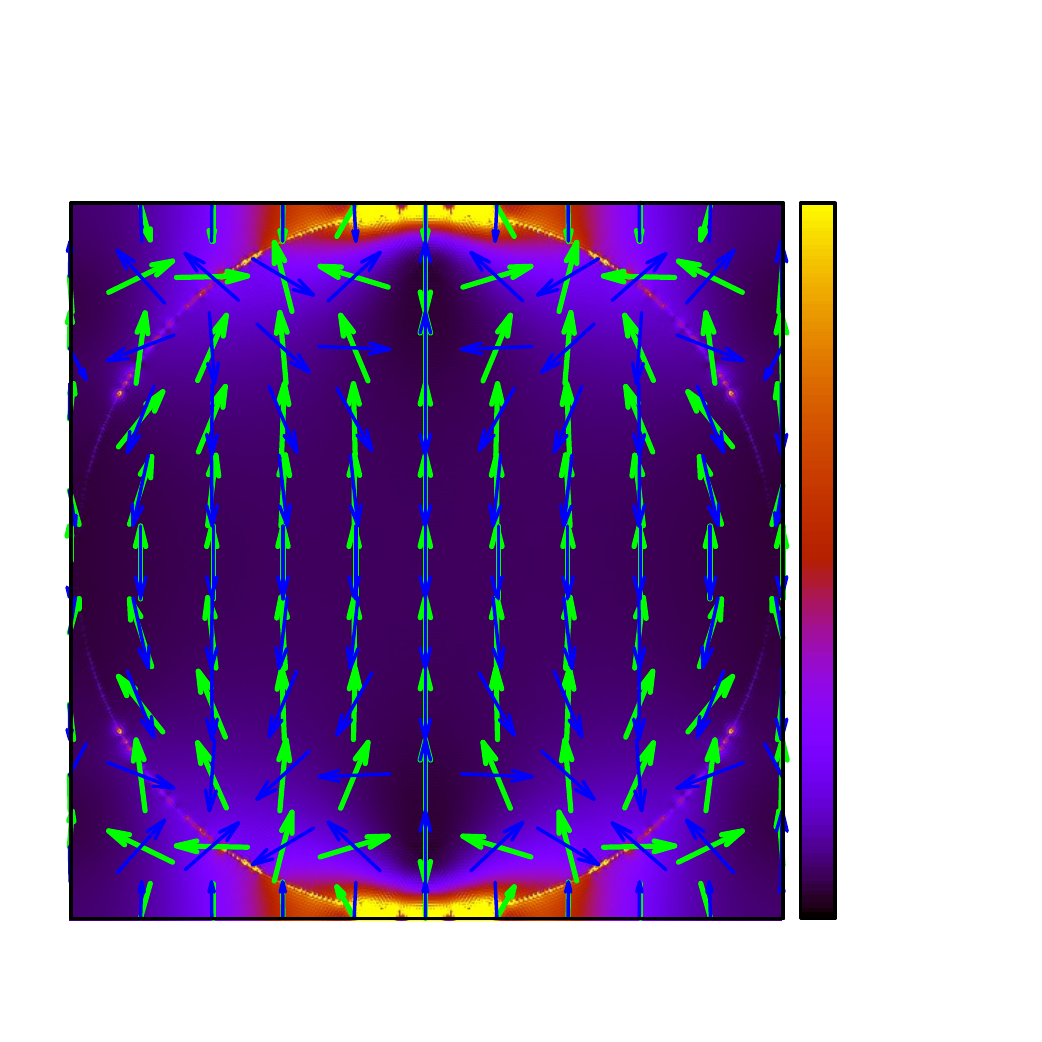}
  \includegraphics[trim = 25 30 50 70,width=0.3\textwidth]
                  {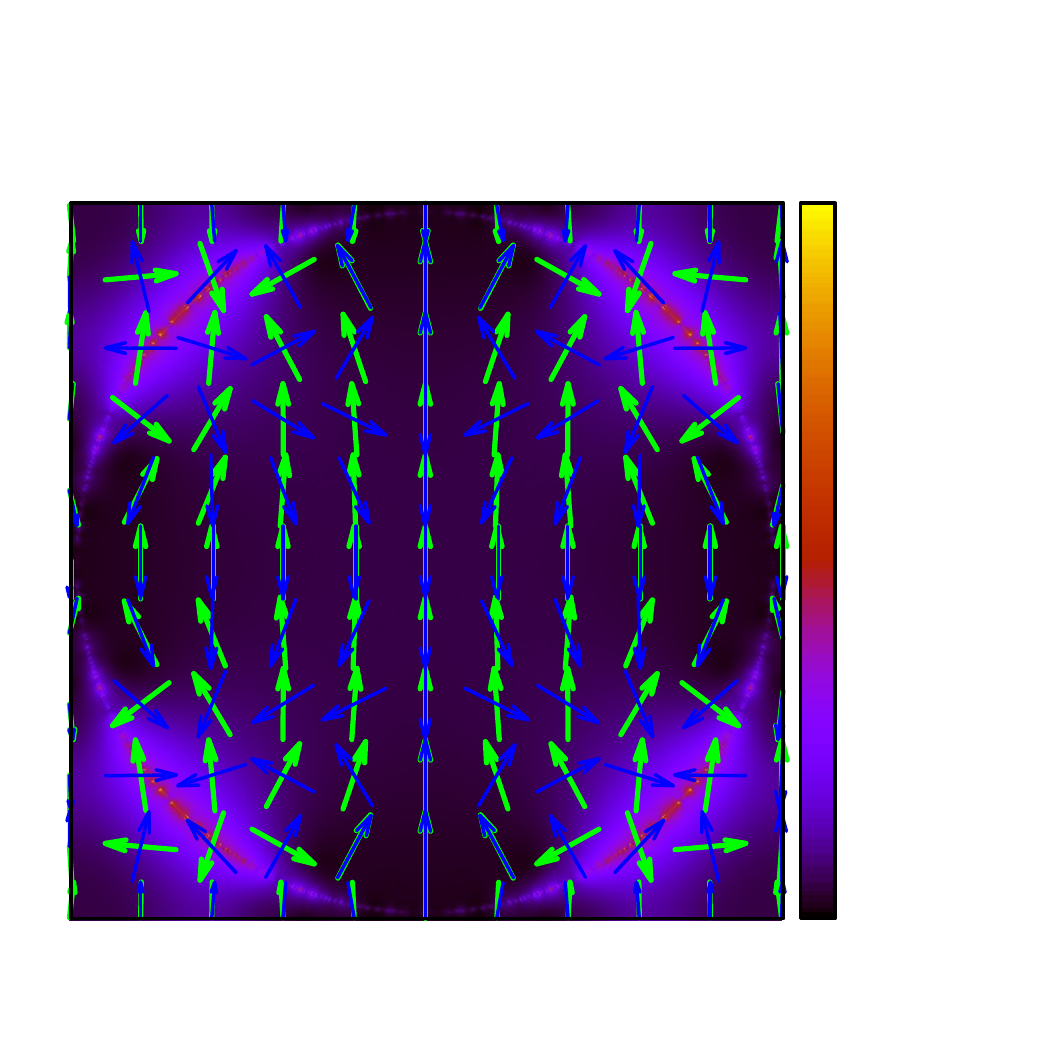}
  \includegraphics[trim = 24 30 50 70,width=0.3\textwidth]
                  {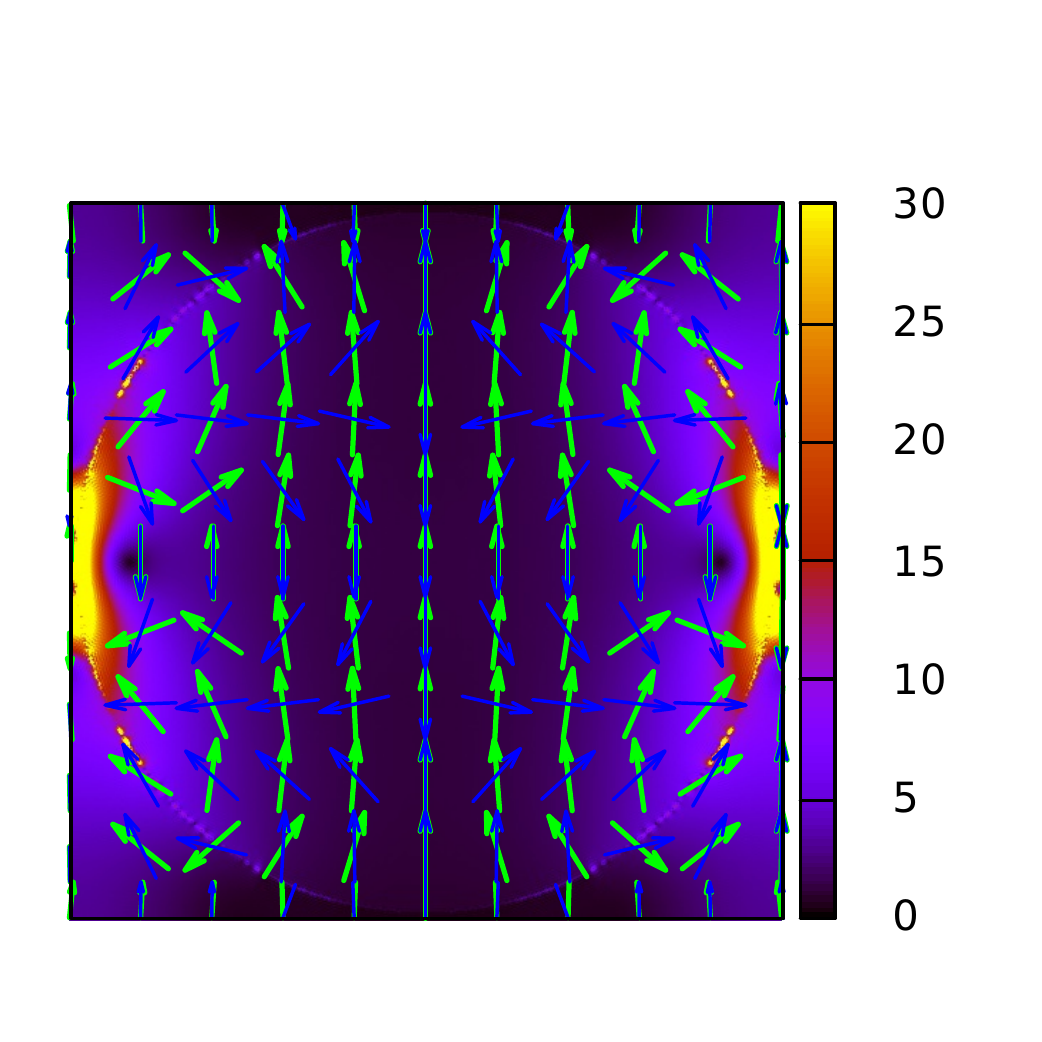}
  $$
  \caption{\label{fig:field_h_air} Direction of the real (green
    arrows) and imaginary (blue arrows) parts of the microscopic field
    and field magnitude (color map) for a square lattice of wires
    with a filling fraction $f=0.75$ as in Fig. \ref{fig:ConcentOrd}
    excited with a homogeneous external field of unit magnitude along
    the vertical direction corresponding from left to right to the
    frequencies $\omega=0.59\omega_p$, $\omega=0.71\omega_p$ and
    $\omega=0.82\omega_p$.
  }
\end{figure}

In Fig. \ref{fig:field_h_air}
we show the microscopic electric field in a lattice of wires, as in
Fig. \ref{fig:ConcentOrd} for three frequencies: the resonance at
$\omega\approx0.59\omega_p$, the dipolar plasmon frequency
$\omega=\omega_p/\sqrt2$ of an individual wire and the resonance at
$\omega=0.82\omega_p$. The calculation was performed with the {\em
  Photonic} code\cite{Mochan(2016)}. Note that in the middle panel the
intensity of the field is the same along the horizontal and vertical
directions. In the left panel, corresponding to a resonance that has
been red shifted from that of the single wire, the field is much more
intense close to the surface of the wires along the vertical
direction, which coincides with the direction of the external field,
while in the right panel, corresponding to a resonance that has been
blue shifted with respect to that of an isolated wire, the intensity
is higher along the horizontal direction, perpendicular to the
external field. We have verified a similar behavior for the other
resonances to the left and right of the isolated surface plasmon.

\begin{figure}
  $$
  \includegraphics[trim =24 30 50
    70,width=0.3\textwidth]{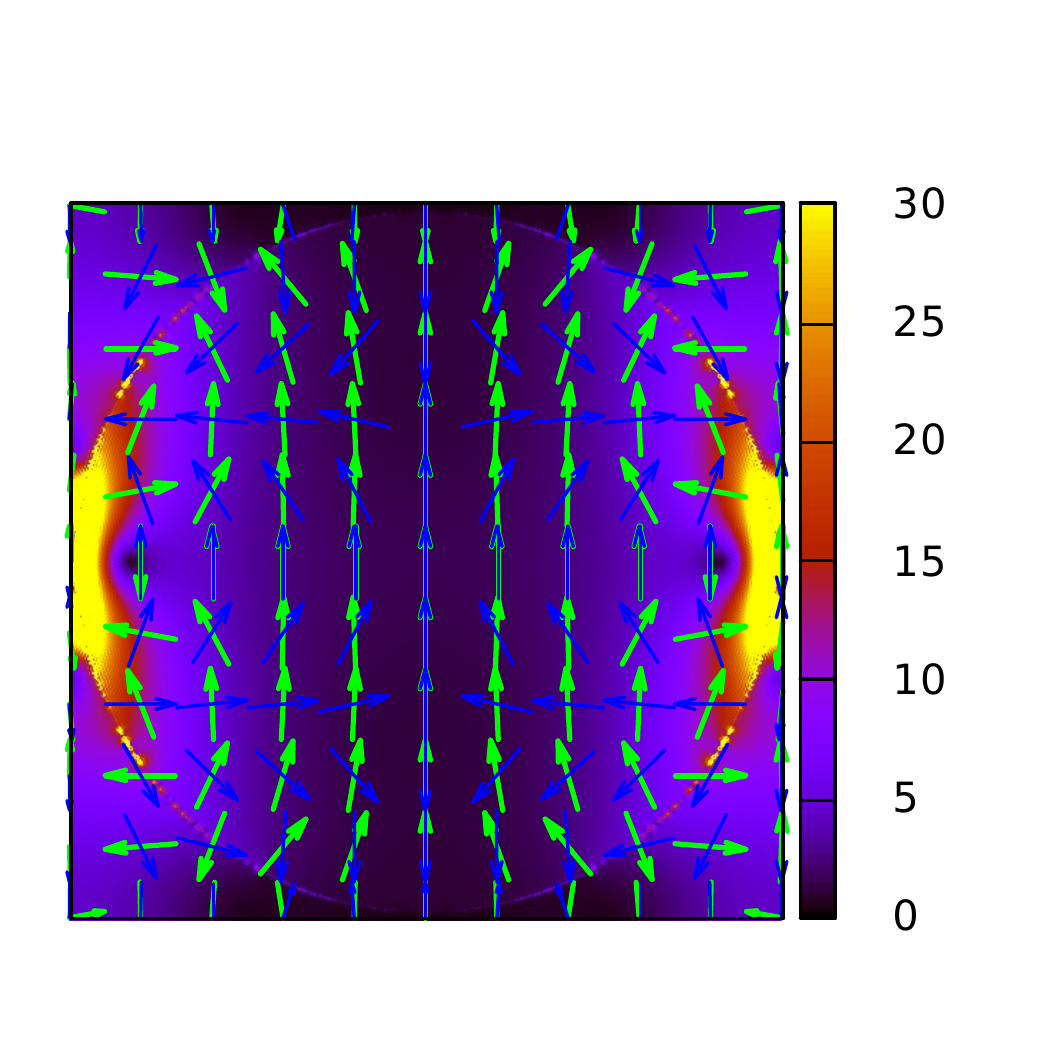}
  \includegraphics[trim =24 30 50
    70,width=0.3\textwidth]{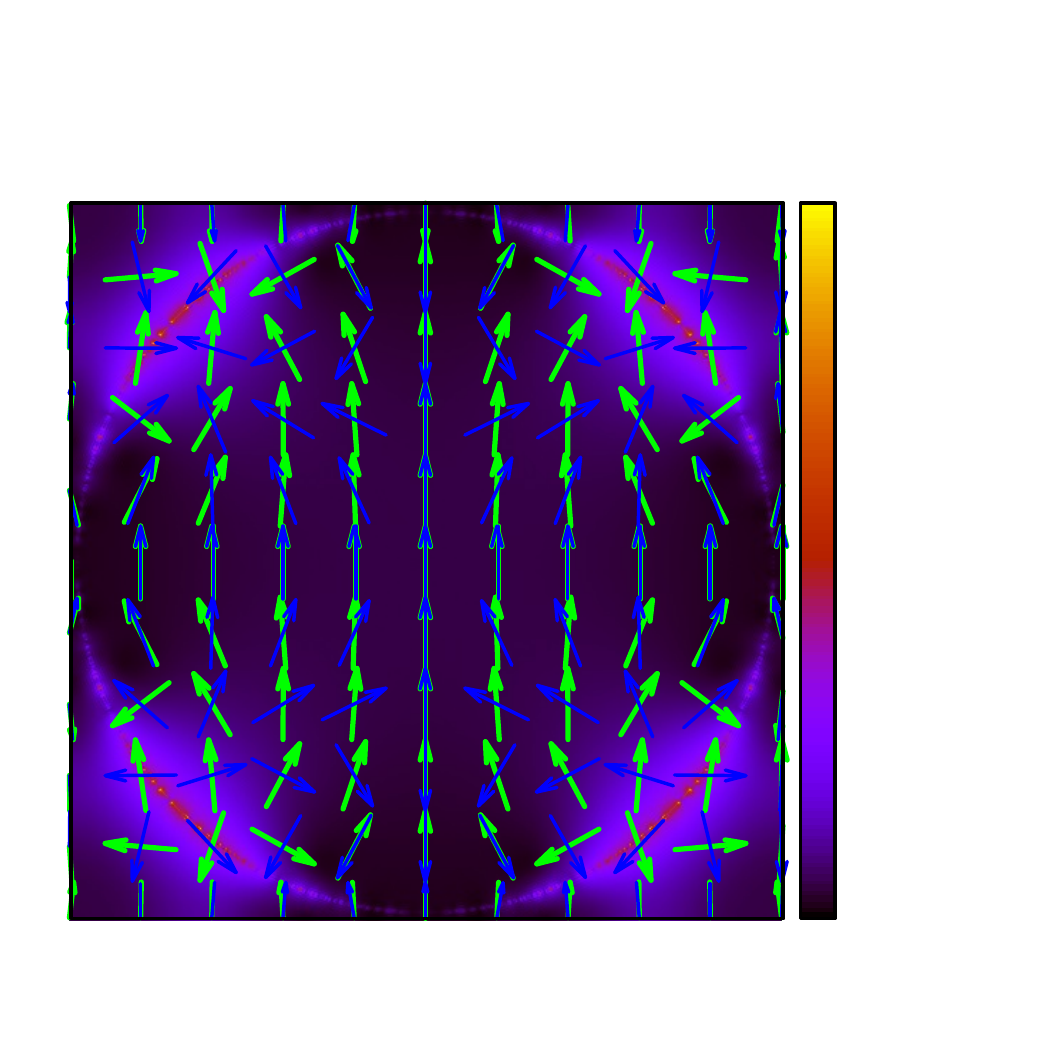}
  \includegraphics[trim =24 30 50
    70,width=0.3\textwidth]{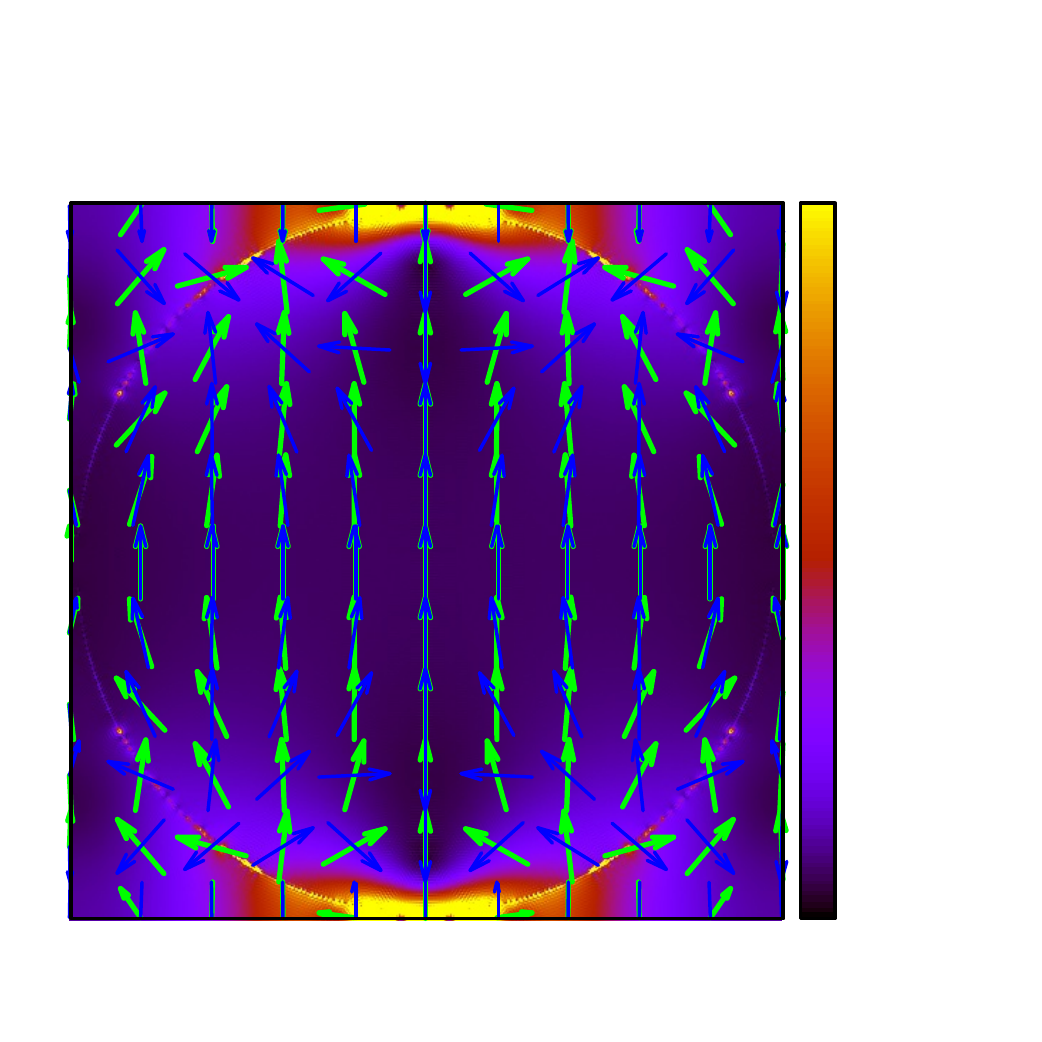}
  $$
  \caption{\label{fig:field_h_metal}
    Direction of the real (green
    arrows) and imaginary (blue arrows) parts of the microscopic field
    and field magnitude (color map) for the system reciprocal to that
    of Fig. \ref{fig:field_h_air}, i.e.,  a square lattice of holes
    within a conductor with a filling fraction $f=0.75$
    excited with a homogeneous external field of unit magnitude along
    the vertical direction corresponding from left to right to the
    frequencies $\omega=0.57\omega_p$, $\omega=0.71\omega_p$ and
    $\omega=0.81\omega_p$.}
\end{figure}

Fig. \ref{fig:field_h_metal} we show the microscopic field for a
lattice of holes, the reciprocal system to that in
Fig. \ref{fig:field_h_air}, 
for the resonance at $\tilde \omega=0.57\omega_p$, the dipolar
surface plasmon frequency $\tilde\omega^{(2)}=0.71\omega_p$  of an
isolated cylindrical hole within a Drude conductor, and
$\tilde\omega=0.81 \omega_p$. We note that the field distribution for
each panel is similar to the field distribution shown in
Fig. \ref{fig:field_h_air} for the corresponding paired frequency
$\omega$, with $\omega^2+\tilde\omega^2=\omega_p^2$. Thus, the panels
of Fig. \ref{fig:field_h_metal} going from left to right correspond to
the panels of Fig. \ref{fig:field_h_air} going from right to left. For
frequencies smaller than that of the isolated surface plasmon the
field is maximum at the surface of the holes in direction normal to
the external field, while at frequencies larger than that of the isolated
surface plasmon the maxima lie along the direction of the external
field. 

\begin{figure}
  $$
  \includegraphics[width=\textwidth]{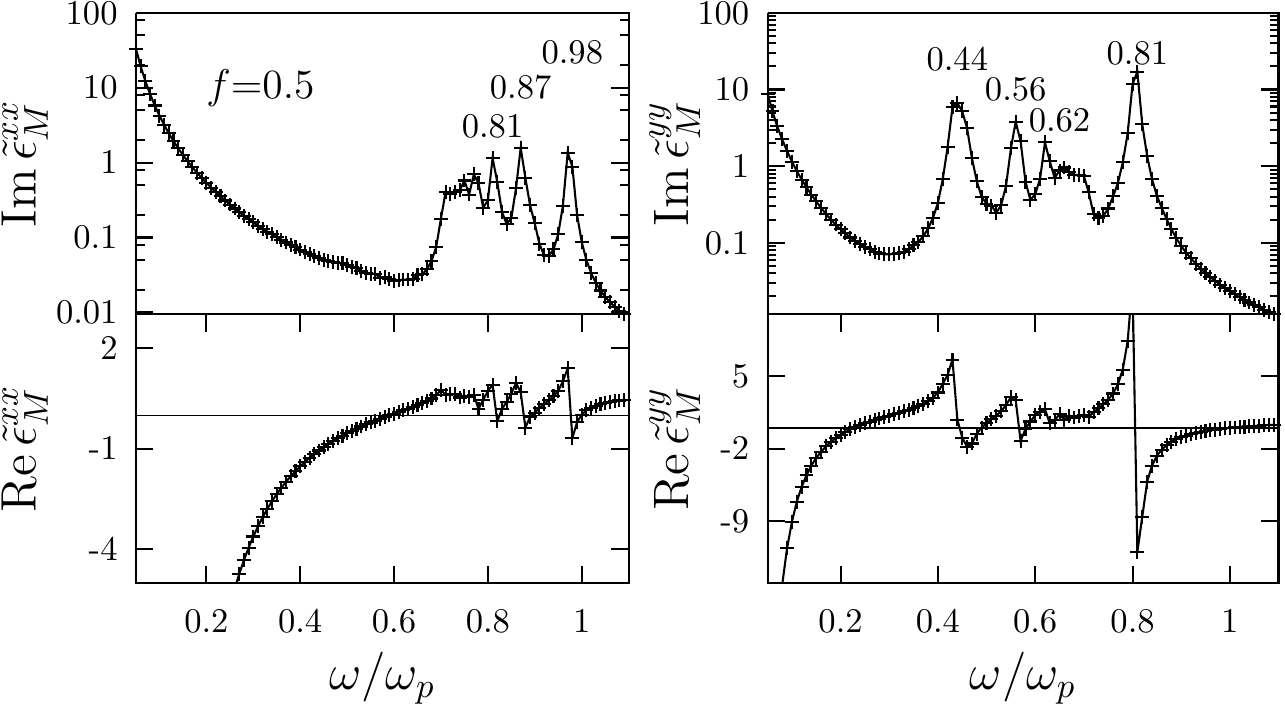}
  $$
  \caption{\label{fig:aniso}Real and imaginary parts of principal
    values of the
    macroscopic response of a 
    rectangular array of cylindrical holes within a Drude conductor as
    a function of frequency for an aspect ratio 3:2 and for a filling
    fraction $f=0.5$, calculated 
    with the HRM (solid) and applying Keller's theorem to the response
    of the corresponding array of cylindrical wires (crosses). The
    field points either along the short side (left panel) or along the
    long side (right panel) of the rectangular unit cell, for the case
    of the array of holes, and in the perpendicular direction for the
    case of wires. } 
\end{figure}

The results above (Figs. \ref{fig:DiluOrd}-\ref{fig:field_h_metal})
were calculated for an isotropic material, for which the Haydock
coefficients, and thus the function $F$ of Eq. (\ref{Haydock}),  are
invariant under rotations. Thus the only change in going
from the system of wires to the system of holes is the substitution
$u\to\tilde u=1-u$. This is not the case for an anisotropic system. In
Fig. \ref{fig:aniso} we show the response of an array of holes calculated
with the HRM and that obtained by applying Keller's theorem
to the response of the corresponding array of wires, as in
Fig. \ref{fig:DiluOrd}, but for an anisotropic rectangular array with
sides in a 3:2 ratio and for a high filling fraction
$f=0.5$. 
The direction of the field for the array of holes was taken along the
short and along the long sides of the rectangular unit cell (left and
right panels respectively).
Note that, unlike Fig. \ref{fig:ConcentOrd}, Fig. \ref{fig:aniso}
shows more resonances shifted towards one side than towards the
opposite side of the surface plasmon of the isolated hole. This is
consistent with Fig. \ref{fig:field_h_metal} which shows that for
$\omega<\omega_p/\sqrt2$ the field is maximum along
the direction normal to that 
of the external field. Thus, it points along the long side of
the unit cell in the left side of the left panel of
Fig. \ref{fig:aniso}, producing no 
visible structure, and along its short side for the right panel,
producing a strong interaction among the holes and thus a rich
resonant structure. On the other hand, for $\omega>\omega_p/\sqrt2$, the
field is stronger along the field's direction, and therefore, it
produces strong 
interactions and a rich structure in the right side of the left panel of
Fig. \ref{fig:aniso} and only a single blueshifted resonance in the right
panel. 
In this case, the Haydock coefficients used for the direct calculation of
the array of holes is different from those used for the array of
wires, due to the $\pi/2$ rotation required by Keller's
theorem. Nevertheless, the direct calculation of
$\tilde\epsilon_M$ and the calculation using Keller's theorem are in
excellent agreement.

\subsection{Disordered Systems}

\begin{figure}
  $$
  \includegraphics[width=0.6\textwidth]{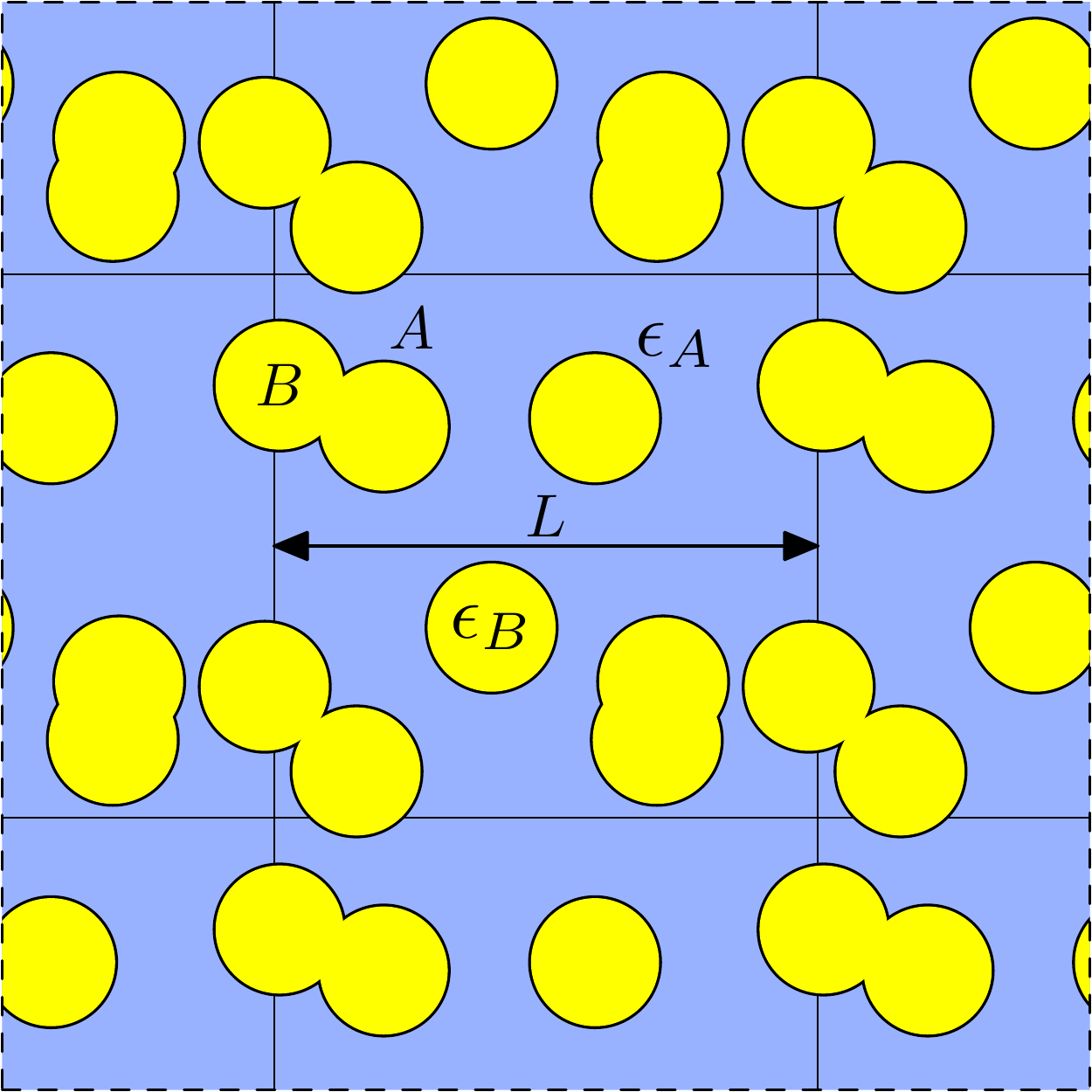}
  $$
  \caption{\label{fig:esquema} Illustration of a disordered system
    approximated by the periodic repetition of a relatively large unit
    cell within which numerous wires occupy random positions.}
\end{figure}
\begin{figure}
  $$
  \includegraphics[width=\textwidth]{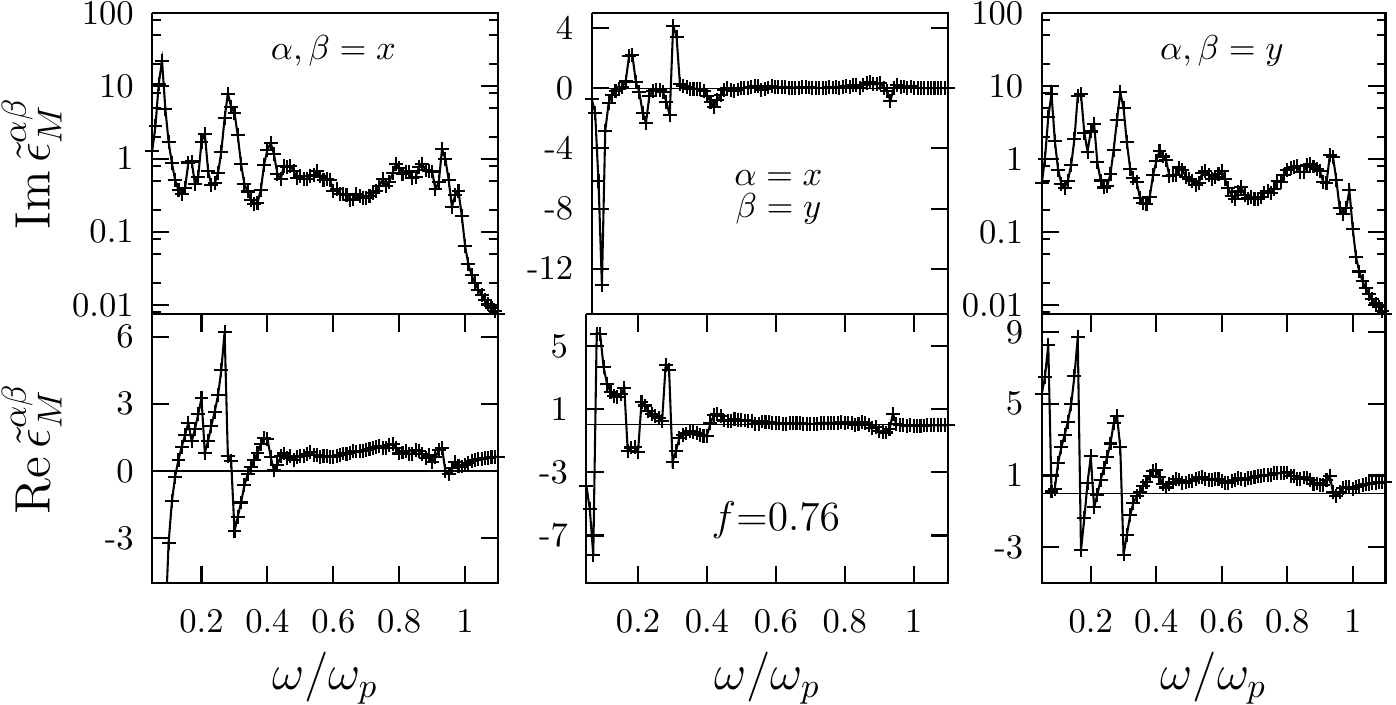}
  $$
  \caption{\label{fig:epsM2DcircleRand} Frequency dependence of the
    real and imaginary parts of the components
    $\tilde\epsilon_M^{\alpha\beta}$ (solid) of the dielectric 
    tensor calculated with the HRM for a single member of an ensemble
    that approximates a 
    disordered system made up of cylindrical holes within a Drude
    conducting host as illustrated in Fig. \ref{fig:esquema}. The
    system consists of 30 cylinders of radius $a=0.12L$ randomly
    distributed without correlation among their positions within a
    square unit cell of side 
    $L$ discretized to $501\times501$ pixels. The 
    filling fraction is $f=0.76$.
    We also show the corresponding result obtained from the dielectric
    tensor of the corresponding system of conducting wires in vacuum
    by employing the tensorial
    version of Keller's theorem (Eq. \ref{KellerF}) (crosses).
  }
\end{figure}

We consider now the response of a disordered system, approximated by
an ensamble of periodic systems with a large unit cell within which
$N$ wires are set at random positions, as illustrated in
Fig.\ref{fig:esquema}. 
In  Fig. \ref{fig:epsM2DcircleRand} we show
the components $\tilde\epsilon_M^{\alpha\beta}$ of the  dielectric
tensor calculated with the HRM for one realization of the 
reciprocal system, consisting of a 
disordered array of cylindrical holes within a conductor. We took
$N=30$ holes and distributed them randomly without correlation,
allowing the holes to overlap. In the same figure we show the result
obtained by first calculating the response
$\epsilon_M^{\alpha\beta}$ of the corresponding disordered system of
conducting wires in vacuum and then using the tensorial version of
Keller's theorem Eq. (\ref{KellerF}). Notice that although the
disordered system is isotropic, a single member of the ensemble with a
finite number of particles is anisotropic, its principal directions
are not necessarily aligned with the cartesian axes and thus they may
depend on frequency, so that $\boldsymbol\epsilon^M$ is not a diagonal
matrix. The response shows a very rich structure due to the strong
coupling between neighbor holes, with fluctuating nearest neighbor
distances and with several pairs of overlapping
neighbors. Nevertheless, Keller's theorem seems to be hold quite well
by our HRM calculations. 

\begin{figure}
  $$
  \includegraphics[width=\textwidth]{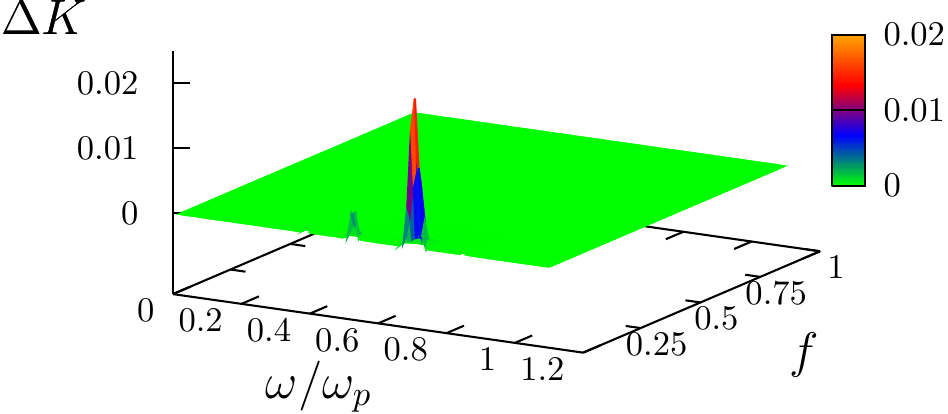}
  $$
  \caption{\label{fig:deltaK} $\Delta K$ as function of frequency
    $\omega$ and filling fraction $f$ for the same realization of the
    disordered system as in Fig. \ref{fig:epsM2DcircleRand}
    but calculated in a unit cell of only $201\times201$ pixels.}
\end{figure}

To explore the fullfilment of Keller's theorem for disordered systems
with different filling fractions, we have varied the radius of the
wires/holes for the same ensemble member as in Fig. 
\ref{fig:epsM2DcircleRand} and we evaluated the deviation from
Keller's theorem  
\begin{equation}\label{DeltaK}
  \Delta K =
  2\left | \frac{\det(\tilde\epsilon^M)\det(\epsilon^M) -
    \epsilon_A^2\epsilon_B^2} {\det(\tilde\epsilon^M)\det(\epsilon^M)
    + \epsilon_A^2\epsilon_B^2} \right|.  
\end{equation}
In Fig.\ref{fig:deltaK} we show $\Delta K$ as function of 
frequency and filling fraction $f$. For this calculation we used a
more modest discretization of only $201\times201$
pixels. Nevertheless, the deviation away from Keller's theorem is very
small except for a few resonances at the smallest filling fractions
$f=0.1$ for which the pixelated representation of the wires and holes
is inadequate.

We have veryfied that our HRM calculations for this system also hold
after averaging over a large enough ensemble.

\subsection{Convergence Acceleration}

Keller's theorem must hold for the {\em exact} nonretarded macroscopic
dielectric tensors 
$\boldsymbol\epsilon_M^E$ and $\tilde{\boldsymbol\epsilon}_M^E$
of a binary 2D composite and its reciprocal system, but it may well
fail for the dielectric tensors $\boldsymbol\epsilon_M$ and
$\tilde{\boldsymbol\epsilon}_M$ obtained from an approximate numerical
calculation.  If we write the exact 
dielectric tensor of a system and its reciprocal as
\begin{equation}\label{exact}
  \boldsymbol\epsilon_M^E = \boldsymbol\epsilon_M+\delta\boldsymbol\epsilon_M,
  \quad
  \tilde{\boldsymbol\epsilon}_M^E = \tilde{\boldsymbol\epsilon}_M +
  \delta\tilde{\boldsymbol\epsilon}_M,
\end{equation}
we can write
Eq.(\ref{KellerF}), as
\begin{equation}\label{ec:keller+delta}
(\boldsymbol\epsilon_M+ \delta\boldsymbol\epsilon_M)
  \mathbf R(\tilde{\boldsymbol\epsilon}_M+ 
  \delta\tilde{\boldsymbol\epsilon}_M)\mathbf R^t=\epsilon_A\epsilon_B,
\end{equation}
that linearizing in $\delta\boldsymbol\epsilon_M$ and
$\delta\tilde{\boldsymbol\epsilon}_M$  becomes a system of four equations
\begin{eqnarray}
  \left(\begin{array}{cc}\delta\epsilon_M^{xx}&\delta\epsilon_M^{xy}\\\delta\epsilon_M^{yx}&\delta\epsilon_M^{yy}\end{array}\right)
  \left(\begin{array}{cc}\epsilon_M^{yy}&-\epsilon_M^{yx}\\-\epsilon_M^{xy}&\epsilon_M^{xx}\end{array}\right)
  + 
  \left(\begin{array}{cc}\epsilon_M^{xx}&\epsilon_M^{xy}\\\epsilon_M^{yx}&\epsilon_M^{yy}\end{array}\right)
  \left(\begin{array}{cc}\delta\epsilon_M^{yy}&-\delta\epsilon_M^{yx}\\-\delta\epsilon_M^{xy}&\delta\epsilon_M^{xx}\end{array}\right)
  =\nonumber\\
  \label{linearized}
  \left(\begin{array}{cc}\epsilon_A\epsilon_B&0\\0&\epsilon_A\epsilon_B\end{array}\right)-
    \left(\begin{array}{cc}\epsilon_M^{xx}&\epsilon_M^{xy}\\\epsilon_M^{yx}&\epsilon_M^{yy}\end{array}\right)
    \left(\begin{array}{cc}\epsilon_M^{yy}&-\epsilon_M^{yx}\\-\epsilon_M^{xy}&\epsilon_M^{xx}\end{array}\right)
\end{eqnarray}
in the six complex unknowns
$\delta\epsilon_M^{xx}$, $\delta\epsilon_M^{xy}=\delta\epsilon_M^{yx}$,
$\delta\epsilon_M^{yy}$,  $\delta\tilde\epsilon_M^{xx}$,
$\delta\tilde\epsilon_M^{xy}=\delta\tilde\epsilon_M^{yx}$, and
$\delta\tilde\epsilon_M^{yy}$,  
which we write as the matrix equation
\begin{equation}\label{MI}
  \mathbf M \mathbf I =\mathbf D,
\end{equation}
where
\begin{equation}\label{I}
  \mathbf I=\left(\begin{array}{c}
    \delta\epsilon_M^{xx}\\\delta\epsilon_M^{xy}\\\delta\epsilon_M^{yy}\\\delta\tilde\epsilon_M^{xx}\\
    \delta\tilde\epsilon_M^{xy}\\\delta\tilde\epsilon_M^{yy}\\
  \end{array}\right),
\end{equation}
and
\begin{equation}\label{M}
  \mathbf M=\left(\begin{array}{cccccc}
    \tilde\epsilon_M^{yy}&-\tilde\epsilon_M^{xy}&0&0&-\epsilon_M^{xy}&\epsilon_M^{xx}\\
    -\tilde\epsilon_M^{yx}&\tilde\epsilon_M^{xx}& 0&\epsilon_M^{xy}&-\epsilon_M^{xx}&0\\
    0&-\tilde\epsilon_M^{yx}&\tilde\epsilon_M^{xx}&\epsilon_M^{yy}&-\epsilon_M^{yx}&0\\
    0&\tilde\epsilon_M^{yy}&-\tilde\epsilon_M^{xy}&0&-\epsilon_M^{yy}&\epsilon_M^{yx}\\    
  \end{array}\right),
\end{equation}
and
\begin{equation}\label{D}
  \mathbf D=\left(\begin{array}{c}
    \epsilon_A\epsilon_B -\epsilon_M^{xx}\tilde\epsilon_M^{yy}+\epsilon_M^{xy}\tilde\epsilon_M^{xy}\\
    \epsilon_M^{xx}\tilde\epsilon_M^{yx}-\epsilon_M^{xy}\tilde\epsilon_M^{xx}\\
    \epsilon_M^A\epsilon_B -\epsilon_M^{yy}\tilde\epsilon_M^{xx}+\epsilon_M^{yx}\tilde\epsilon_M^{yx}\\
    \epsilon_M^{yy}\tilde\epsilon_M^{xy}-\epsilon_M^{yx}\tilde\epsilon_M^{yy}\\
  \end{array}\right).
\end{equation}
Although Eq. (\ref{MI}) is underdetermined and doesn't have a unique
solution, one may attempt to obtain the {\em smallest} corrections
$\delta\boldsymbol\epsilon_M$ and
$\delta\tilde{\boldsymbol\epsilon}_M$ that when added to the
approximate results $\boldsymbol\epsilon_M$ and
$\tilde{\boldsymbol\epsilon}_M$ yield response functions that better
fulfill Keller's theorem and that may thus be expected to better
approximate the exact results. To that end we perform a singular value
decomposition (SVD) \cite{Press(1992)}
\begin{equation}\label{SVD}
  \mathbf M = \mathbf U \boldsymbol\Sigma \mathbf V^t,
\end{equation}
where $\mathbf U$ and $\mathbf V$ are column-orthogonal matrices and
$\boldsymbol\Sigma$ is a 
diagonal matrix, obtaining
\begin{equation}\label{ec:I}
  \mathbf I= \mathbf V \boldsymbol \Sigma^{-1} \mathbf U^t \mathbf D. 
\end{equation}

To illustrate the use of Keller's theorem to improve the convergence
of numerical calculations we use the HRM to make crude calculations of
$\epsilon_M$ and $\tilde\epsilon_M$ and then correct our calculations
using the procedure above get better approximations for which the
deviations from Keller's condition is smaller.

\begin{figure}
  $$
  \includegraphics[width=0.8\textwidth]{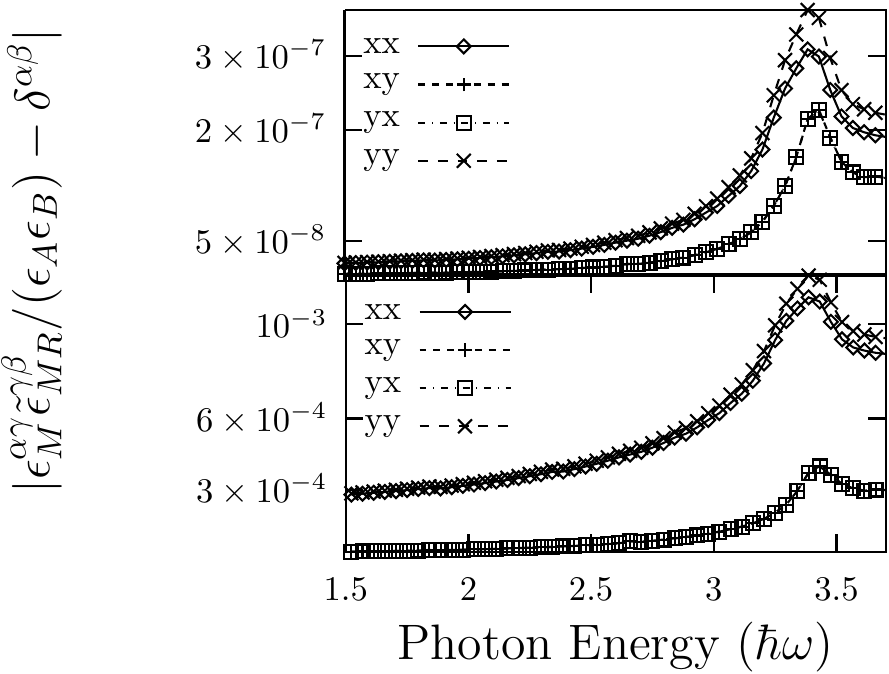}
  $$
  \caption{\label{fig:AirSiRandcbFix} Absolute value of the different
    components of the departure of the computed dielectric tensors
    $\boldsymbol\epsilon_M$ and
    $\tilde{\boldsymbol\epsilon}_M$ from Keller's theorem
    (\ref{KellerF}) for an ensemble of one hundred realizations of
    a random checkerboard with ten thousand particles each consisting of Si
    prisms within vacuum. The bottom panel panel shows the result of
    the HRM calculation and the upper panel the result after adding
    corrections from Eq. (\ref{ec:I}).}
\end{figure}

In Fig. \ref{fig:AirSiRandcbFix} we show the deviation from Keller's
condition for a system made up of square Si prisms randomly occupying
the sites of a square array within vacuum, with a filling fraction
$f=1/2$. 
We see that adding the correction (\ref{ec:I}) diminishes the
deviation from Keller's condition by more than four orders of
magnitude, from the order of $10^{-3}$ to $10^{-7}$ or better. We
remark that this is an isotropic system symmetrical under the
interchange of components for which the exact dielectric response is
completely determined by Keller's theorem through
Eq. (\ref{symmetric}).

\begin{figure}
  $$
  \includegraphics[width=0.8\textwidth]{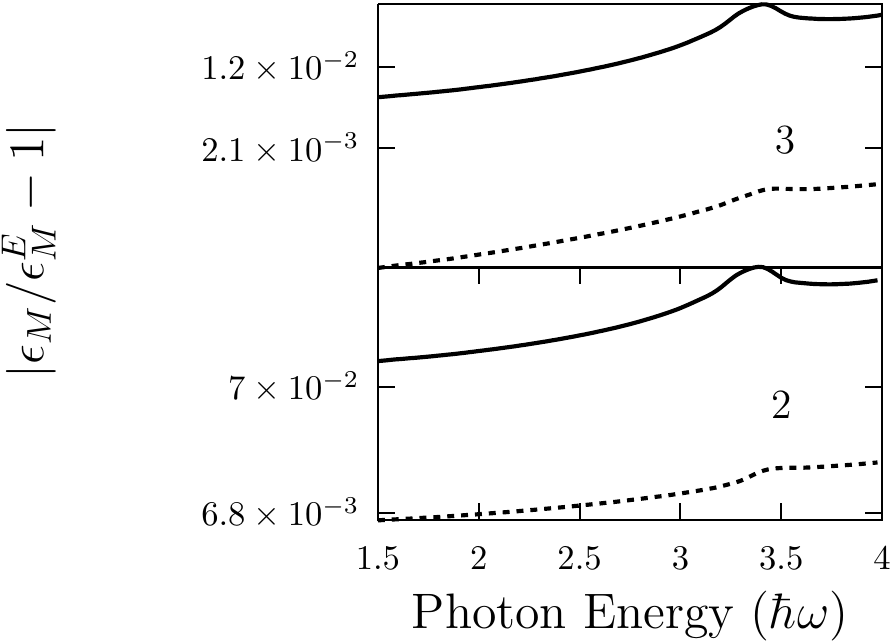}
  $$
  \caption{\label{fig:AirSiMilton} Relative difference between the
    dielectric function  of $\epsilon_M$ calculated with the HRM for a
    square array of square Si prisms with filling fraction $f=1/4$
    within vacuum, and the exact response (\ref{mortola}) before
    (solid lines) and after (dashed lines) the correction (\ref{ec:I})
    is applied. The HRM calculations were performed with an extremely
    small number of Haydock iterations: 2 for the lower panel and 3
    for the upper panel. 
  }
\end{figure}

Now we turn our attention to a system proposed by Mortola and Steff\'e
\cite{Mortola(1985)} consisting of a square array of square prisms
with filling fraction $f=1/4$. It turns out that this system has the
exact solution \cite{Milton(2001)}
\begin{equation}\label{mortola}
  \epsilon_M^E = \epsilon_A \sqrt{
    \frac{\epsilon_A+3\epsilon_B}{\epsilon_B+3\epsilon_A} }.  
\end{equation}
It has been shown \cite{Mendoza(2016)} that the HRM is capable of
reproducing numerically this results, even for metallic phases. 
In Fig. \ref{fig:AirSiMilton} we display the relative
error of the numerical calculation of the macroscopic response of a
systems made up of a square lattice of square Si prisms with a filling
fraction of f=1/4 calculated with an extremely small number of Haydock
pair of coefficients $n=2$ and $n=3$. Not surprisingly, the crude
numerical results have a large discrepancy of a few percent from the
exact result. Nevertheless, an order of magnitude accuracy increase is
obtained by applying the correction (\ref{ec:I}). Furthermore, better
inicial results benefit from even higher accuracy increases.

\section{Conclusions}\label{conclusions}
We obtained a version of Keller's theorem relating the macroscopic
dielectric tensor $\boldsymbol\epsilon_M$ of 2D binary composite
systems to the corresponding response  $\tilde{\boldsymbol\epsilon}_M$
of their reciprocal systems, with the same
geometry but with their two components interchanged. The derivation
assumes that the texture of the system has a lengthscale that is much
smaller than the wavelength of light, but otherwise, is valid for
finite frequencies and may be applied to dispersive and dissipative
materials. We obtained results
for the generic anisotropic case, and special results for the
isotropic case and for systems symmetric under interchange of
materials. Our results are 
based on a general homogenization procedure that does not require the
fields to be irrotational or solenoidal, as we make no assumption
about the absence of sources in the derivation. Although
Keller's theorem is frequently stated in terms of the electrical
conductivity $\boldsymbol\sigma_M$, we show that in general this
response only obeys Keller's theorem in the limit of very low
frequencies. Nevertheless, we obtained a generalization of Keller's
theorem for the 
conductivity at finite frequencies. We developed a few applications of
Keller's theorem. Thus we showed that the expression for the response
of a 1D superlattice perpendicular to its axis is determined by its
response along its axis. We verified that common effective medium
theories, such as Maxwell Garnett's and Brugemman's expressions, do
obey Keller's theorem. We showed how one may employ Keller's theorem
to check the accuracy of numerical computations, we showed that for
each resonance of an isotropic system there is a corresponding
resonance of the reciprocal system described by a corresponding
spectral variable and 
with the same microscopic field distribution. We illustrated the use
of Keller's theorem to test model calculations for ordered,
disordered, isotropic, and anisotropic systems. Finally, we showed that
Keller's theorem may be employed to increase the accuracy of
approximate numerical calculations.

\section*{Acknowledgments}
We acknowledge support from DGAPA-UNAM under grant IN113016, from
FONCYT under grant PICT-2013-0696, and from SGCyT-UNNE under grant
PI-F008-2014.
\appendix
\section{Appendix: Anisotropic materials}

If the materials $A$ and $B$ were themselves anisotropic, then
instead of Eq. (\ref{epsB}) we would have
\begin{equation}\label{epsTen}
    \boldsymbol\epsilon(\vec r)=\boldsymbol\epsilon_A (1-B(\vec r)) +
    \boldsymbol\epsilon_B B(\vec r)=\epsilon_A \mathbf U_A (1-B(\vec r)) +
    \epsilon_B \mathbf U_B B(\vec r),
\end{equation}
where $\boldsymbol\epsilon_\alpha\equiv\epsilon_\alpha\mathbf
U_\alpha,$ ($\alpha=A,B$),  
$\epsilon_\alpha\equiv\sqrt{\det{\boldsymbol\epsilon_\alpha}}$ and $\mathbf
U_\alpha$ is a unimodular matrix, $\det\mathbf U_\alpha=1$. Then,
\begin{equation}\label{epsTenInv}
  \boldsymbol\epsilon^{-1}
  =\frac{\tilde{\boldsymbol\epsilon}}{\epsilon_A\epsilon_B}, 
\end{equation}
where
\begin{equation}\label{epsTenTilde}
    \tilde{\boldsymbol\epsilon}=  \mathbf R(\epsilon_B \mathbf U_A^t
    (1-B(\vec r)) + \epsilon_A \mathbf U_B^t B(\vec r)) \mathbf R^t
\end{equation}
is the response of the reciprocal system obtained by interchanging the
scalar responses of $A$ and $B$ and transposing and rotating their
orientation dependence (but without interchanging it). From here, we
can follow all steps of section \ref{teoria} from Eq. (\ref{epsM-3})
to the main 
result (\ref{KellerF}) \cite{Mendelson(1975)}.  The only difference
being the more 
complicated and somewhat artificial definition of the reciprocal
system above. Some simplifications may be done when considering
the symmetric nature of the matrix $\mathbf U_\alpha$, and in the
special case where $\mathbf U_A=\mathbf U_B$.

\bibliography{referencias}
\end{document}